# Type and Degree of Covalence:
# Empirical Derivation and Implications

**Yevgeny Rakita[1] \*, Thomas Kirchartz[2,3], Gary Hodes[1], David Cahen[1]**

\* [yevgev@gmail.com](mailto:yevgev@gmail.com)

*Current address: Applied Physics and Applied Mathematics, Columbia University, 500 w 120st Mudd #1105, 10025, New York, NY, USA ; [yr2369@columbia.edu](mailto:yr2369@columbia.edu)*

1. Department of Materials and Interfaces, Weizmann Institute of Science, Rehovot, 76100, Israel
2. IEK5-Photovoltaics, Forschungszentrum Jülich, 52425 Jülich, Germany
3. Faculty of Engineering and CENIDE, University of Duisburg–Essen, Carl-Benz-Strasse 199, 47057 Duisburg, Germany

**Abstract:**

The 'covalent character' of the bond in a semiconducting material is a feature that is usually approached theoretically and *empirical* quantification of covalence is absent. Covalent bonds are 'classically' referred to as attractive interactions, but it has been proposed that 'anti-bonding' character of valence electrons may be beneficial, by leading to 'defect-tolerance'. We develop an approach to identify both the *type* [i.e., attractive ('bonding') or repulsive ('anti-bonding')] and *degree* of the covalent part in a chemical bond of semiconductors. We argue, and prove, based on *empirical* correlations, that the *relative structural polarizability*, RSP, the ratio between the structural and the electronic (hard-sphere) polarizabilities, measured as $\left(\frac{\varepsilon_{ion}}{\varepsilon_{elect}}\right) \approx \left(\frac{\varepsilon_s}{\varepsilon_\infty} - 1\right)$, is a reliable metric for the nature of the covalent bonding. Also the deformation potential, or bandgap-pressure coefficient, $\frac{\Delta E_g}{\Delta p}$, can give a rough indication of covalent bonding type. We suggest structural and chemical trends, common for semiconductors with 'anti-bonding' valence band, which differ from those of the 'classical' set of semiconductors with 'bonding' valence band. We show that the nature of a covalent bond has significant implications for fundamental (opto)electronic properties in heteropolar compounds. Among more than 40 different compounds, the covalent nature correlates with the electronic effective mass, the screening capability of perturbative electric fields and, as a consequence, the mobility of free charges. These results provide tools to identify the bond nature of semiconducting materials and can serve to guide discovery and engineering of new materials with desired properties.

**Popular Summary:**

The way atoms attach to each other defines the function(s), e.g., mechanical, optical, electronic, of a given material. The nature of the chemical bond is, therefore, one of the most fundamental issues in materials. Both ionic interactions, i.e., resulting from electrical charges associated with the atoms, and covalent ones, i.e., the sharing of electrons between nuclei of different atoms, are usually viewed as forces that attract between atoms to form a rigid structure. Although less common for solid materials, it was shown theoretically to be possible for covalent interactions at the chemically-active electronic shell (or 'valence-band maximum') of semiconductors to reverse their more common nature and become repulsive, i.e., act against bonding.

Some semiconductors with such predicted 'anti-bonding' valence-band maximum levels (such as halide perovskites) show experimentally some amazing (opto-) electronic properties. Predictions that 'anti-bonding' character can allow tolerance for existing defects, at least in part, can explain the superior properties of such semiconductors.

Although there are known experimental ways to estimate the *degree* of the covalent nature (e.g., electronegativity), this was not possible hitherto for the *type*, i.e., distinguishing whether a material exhibits 'bonding' or 'anti-bonding' covalent interactions. We have developed a simple way to reveal the *complete* nature (*both* type and degree) of chemical bonds, using experimental data. After confirming our development with 'classical' models and theoretical predictions, with a set of ~40 different functional semiconductors, we show how knowledge of the *complete* nature of covalent bonding is of critical importance for fundamental properties of semiconductors.

## I.     Introduction:

The bond nature of materials can be related to many of the properties of a material. The concept of a chemical bond was extensively discussed in the scientific literature over the last century[1–3], mostly mentioning metallic, covalent and electrostatic (ionic) bonds. The difference between these three bond types is the degree of delocalization of the valence electrons across the structure: completely delocalized (metallic), shared between atoms (covalent) and completely localized (ionic).

Purely covalent bonds are found in 'homopolar' materials (containing only one element, e.g., Si). Partially covalent bonds exist in 'heteropolar' materials (with different elements; e.g., GaAs, PbSe), and these differ in their relative degrees of 'covalence' and 'ionicity', where 'covalence'=(1-'ionicity'). Although in the literature 'ionicity', a measure of the charge transferred to/from a neighboring atom (leading to electrostatic attraction), is the more common term, we refer to 'covalence' since it can be *attractive* or *repulsive*.

The atomic type, formal oxidation state, structural symmetry, and covalent character of a bond are fundamental features that determine many of the functionally-significant thermal, electronic and mechanical properties of a material. [2,4] The bonding character of a compound, specifically its covalence, is a feature that is usually approached theoretically and a common *empirical* way of assessing it is absent. Covalent bonds are 'classically' referred to as attractive interactions, but it has been argued that 'anti-bonding' covalent character of the valence electrons can lead to interesting properties.[5–7] For example, when considering semiconductors, the interaction of electronic charge carriers with defects usually refers to their densities and interaction cross-section; the latter relates to the charge and the energy level of the defect within the optical bandgap and electron-phonon interaction.[4,8] In favorable cases where both the valence and conduction band extrema are constructed of 'anti-bonding' covalent interactions, as for halide perovskites (HaPs) [(see Endnote i)],[6] Pb-chalcogenides[7] and anti-perovskites ($Cu_3N$)[5], intrinsic imperfections have states with energies in the bands or just in the gap, which then are expected to be electrically and optically benign.

How can we tell, experimentally, if a covalent bond is attractive or repulsive? Here we suggest to distinguish, both qualitatively and quantitatively, the type (attractive 'bonding' or repulsive 'anti-bonding') and degree (less or more 'ionic') of a bond in crystalline solids, using experimentally accessible parameters, namely the relative structural polarizability (*RSP*) (defined below) and the bandgap-pressure dependence, which is directly linked to the *deformation potential*. After correlating RSP with classical definitions for the degree of covalence (estimated from empirical quantities by Pauling and others)[1,2], we show that RSP is a very good parameter to define both type and degree of covalence, while use of the deformation potential is more limited.

To test the practical relevance of our extended empirical approach for defining the bond nature with RSP, we show relations between the covalent bond character and properties related to charge dynamics, i.e., electronic free carrier lifetime, effective mass and mobility. The set of parameters used here, should, in the future be expanded and compared to theory to test the potential of RSP as a figure of merit to predict the functions of more materials, existing and new ones.

## II.  **Model:**

We use a simple model to rationalize the correlations that are presented in the 'Results' section. We consider periodic crystalline systems, like Si, GaAs or LiF, that are represented by the three figures in Figure 1(i) – left to right, respectively. In Figure 1(i), the interatomic bond in a *homopolar* (Si-like) system involves covalent bonding with an energy of $U_{cov}$. In a *heteropolar* (GaAs- or LiF-like) system, in addition to delocalization, charge separation introduces an additional electrostatic energy ($U_{ion}$) to the total bond energy, so $U_{bond} = U_{cov} + U_{ion}$. [(see Endnote ii)] We will focus on materials in which electrons are still delocalized (to some degree),i.e., *not* purely ionic, [(see Endnote iii)] which then can result in semiconducting behaviour. The degree of electronic interference between neighboring atoms will represent the *degree* of covalence in a material.

---

[i] **Ha**lide **P**erovskites (HaP) are semiconductors (SCs) with remarkable optoelectronic quality and, consequently, performance with solar to electric power conversion efficiency of >24%.[9] One feature, which differentiates HaPs from other SCs, is their (very) low defect density, especially when considering the (low) energy input and complexity required for their fabrication. Experimentally-derived values of trap density in HaPs range from ~$10^{10}$ cm$^{-3}$ for solution-grown single crystals and $10^{13}$-$10^{16}$ cm$^{-3}$ for polycrystalline films.[10–13]

Although we will use the term 'covalence', 'ionicity' is the more common term.[14] 'Ionicity' was first defined by Pauling from calorimetric measurements from which electronegativity was derived, and further developed by others.[2,3] Pauling's ionicity term (the later one, which included the concept of 'resonating bonds'), is defined as: [2]

Eq. 1) $$f_i' = 1 - \frac{N}{M}\left(\exp\left[-\frac{(X_A - X_B)^2}{4}\right]\right)$$

Since 'covalence' ≡ (1 – 'ionicity'), 'covalence' can be expressed as:

Eq. 2) $$f_c' = \frac{N}{M}\left(\exp\left[-\frac{(X_A - X_B)^2}{4}\right]\right)$$

Here $X_A$ and $X_B$ are the electronegativity of the different atoms, and $N$ and $M$ are the anion valence and the effective coordination number of the systems. [(see Endnote iv)]

This, as well as any other expression for 'covalence', does not give information on whether the covalent bond is attractive ('bonding') or repulsive ('anti-bonding'). To get empirical insight into both *degree* and *type* of 'covalence' we proceed with a "thought experiment" by introducing a perturbation to Figure 1(i): a mechanical displacement (Figure 1(ii)) or an electric field (Figure 1(iii)). An electric field will force the material to respond by displacing the electric charge (electronic or ionic), which will generate an opposing electric field.

The proportionality factor between the applied electric field strength ($E$) and the electrical displacement field ($D$) is the dielectric function, $\varepsilon_{(\omega)}$: [(see Endnote v)]

---

ii  In GaAs, which is mostly covalent, $U_{cov} > U_{ion}$, while in LiF, which is mostly 'ionic', $U_{cov} < U_{ion}$. As the cohesive energy in GaAs is mostly dominated by covalent bonds, $U_{cov}$ should be a strongly attractive ('bonding'), while since the cohesive energy for LiF is mostly electrostatic, $U_{cov}$ can, in principle, also be 'repulsive'.

iii A 'purely ionic' system is a hypothetical situation since there will always be *some* degree of electron sharing.

iv  $X_i$ is, in principle, an empirical value, derived from of the excess heat of formation (calorimetric measurements) of the A-B structure with respect to that of the elemental bond A-A and B-B. Originally, it was conceived for molecules. In some editions of Pauling's "The Nature of the Chemical Bond" (Cornell Univ. press, 1939 (1st Ed.) by Pauling, 1945, 1960), its use for non-molecular crystals systems was apparently included. As can be seen from the table in Huheey's textbook, *Inorganic Chemistry*,[15] other electronegativity scales exist. What is most relevant here for us is a scale that works well for non-molecular, extended solids. Such an example is presented by Phillips,[16] where we later show that using Pauling's 'resonating-bond' concept (Eq. 2) with the electronegativity values refined by Phillips correlates well with RSP.

v  $[D] = \frac{C}{m^2}$  ;  $[E] = \frac{V}{m}$  ;  $[\varepsilon] = \frac{F}{m} = \frac{C}{V \cdot m}$. Also: the dielectric function, $\varepsilon_{(\omega)} = \varepsilon_0 \cdot \varepsilon_{r(\omega)}$, where $\varepsilon_0$ is the vacuum permittivity and $\varepsilon_{r(\omega)}$ is the frequency-dependent relative permittivity. For convenience, $\varepsilon_{r(\omega)}$ will be used interchangeably with $\varepsilon_{(\omega)}$.

$$\text{Eq. 3)} \qquad D_{(\omega)} = \varepsilon_{(\omega)} \cdot E_{(\omega)}$$

where $\omega$ is the frequency of the applied electric field. Following Figure 1(ii, a displacement), at a displacement ,$\Delta r$, charge density is being changed by $\delta q$, leading to generation of an electric field, $E$. Similarly, an applied external electric field will create $\delta q$, which will then generate a displacement of charge density by $\Delta r$ (see Figure 1(iii)). Unlike the former case, Figure 1(ii), the latter case, Figure 1(iii), does not necessarily involve atomic displacement, but can involve displacement of only the electronic shell. In both cases, a 'polarization energy', $\Delta P_{(\omega)}$, which is equivalent to the electrostatic work, $\Delta U_{ion}(\omega)$, will be generated. Defining the displacement of the charge from equilibrium as: $D_{(\omega)} = \frac{\delta q_{(\omega)}}{\Delta r^2}$, one can represent $\Delta P_{(\omega)}$ as: [see Endnote vi)]

$$\text{Eq. 4)} \qquad \Delta P_{(\omega)} \approx \delta q_{(\omega)} \cdot (\Delta E_{(\omega)} \cdot \Delta r) = D_{(\omega)} \cdot \Delta E_{(\omega)} \cdot \Delta r^3$$

Using Eq. 3, converting $r^3$ to volume, $r^3 \approx V$, and multiplying by the initial volume, $V$, we can write: [see Endnote vii)]

$$\text{Eq. 5)} \qquad \Delta P_{(\omega)} \approx \varepsilon_{(\omega)} \cdot E_{(\omega)}^2 \cdot \Delta V$$

Consequently, we define the polarization energy *per volumetric strain*, $\left(\frac{\Delta V}{V}\right)$:

$$\text{Eq. 6)} \qquad \frac{\Delta P_{(\omega)}}{\left(\frac{V}{\Delta V}\right)} = \Delta \breve{\boldsymbol{P}}_{(\omega)} \approx \boldsymbol{\varepsilon}_{(\omega)} \cdot \boldsymbol{E}_{(\omega)}^2 \cdot \boldsymbol{V}$$

Now we consider the mechanical displacement energy. With $B$ being the bulk modulus, the strain energy, $\Delta U_{disp}$, is often expressed via

$$\text{Eq. 7)} \qquad \Delta U_{disp} = \frac{1}{2} B \cdot \left(\frac{\Delta V}{V}\right)^2 \cdot V.$$

Using the approach of 'energy *per volumetric strain*' (represented with a diacritic ' $\smile$ ' sign), one can write $\Delta \tilde{U}_{disp}$ as:

$$\text{Eq. 8)} \qquad \Delta \tilde{\boldsymbol{U}}_{\boldsymbol{disp}} \approx \boldsymbol{B} \cdot \Delta \boldsymbol{V} .$$

The missing part, $\Delta \tilde{U}_{cov}$, which expresses the change in the hybridization energy as a result of the change in the overlapping orbitals (due to changing distance and geometry), may be extracted from energy conservation, i.e.:

$$\text{Eq. 9)} \qquad \Delta \tilde{U}_{disp} = \Delta \breve{P}_{ion} + \Delta \tilde{U}_{cov}.$$

To some (limited) extent, in solids the change in energy of the valence band maximum (VBM) should also represent $\Delta \tilde{U}_{cov}$. Unlike $\Delta \breve{P}_{ion}$, which will always be attractive upon displacement, $\Delta \tilde{U}_{cov}$ may have

---

either of both characters: *attractive ('bonding')* or *repulsive ('anti-bonding')*. Therefore, the algebraic sign of $\Delta \tilde{U}_{cov}$ provides insight into the type of the covalent bond in addition to its magnitude.

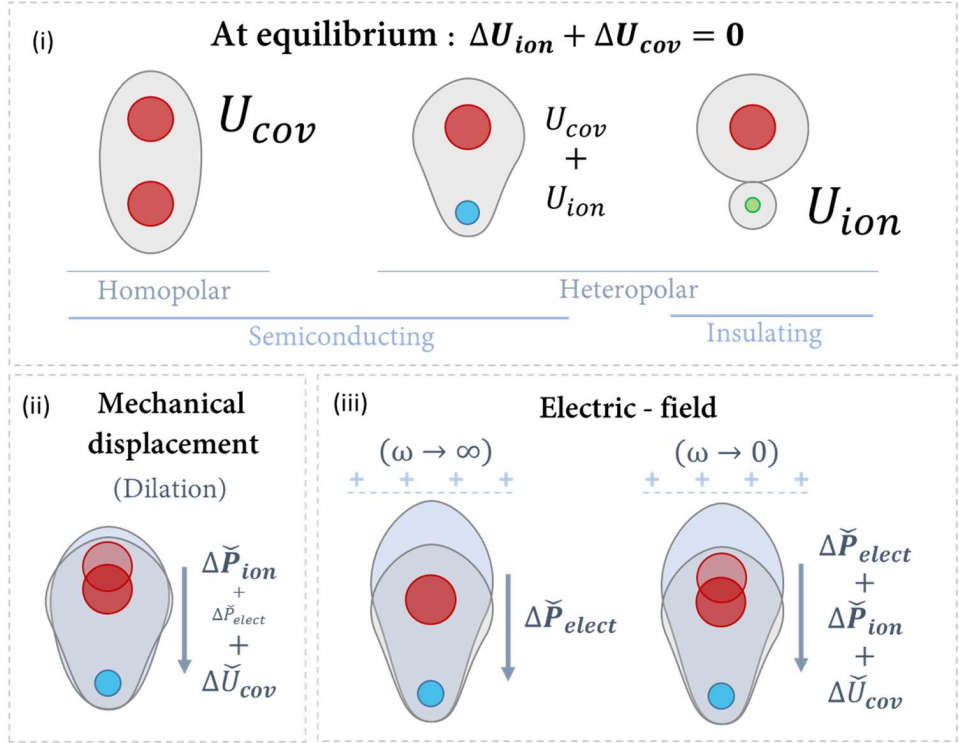

Figure 1: (*i*) Simplistic schematic representations of a single bond in 'homopolar' or 'heteropolar' crystalline systems with changing degree of covalence at equilibrium. The left side represents purely 'covalent' materials (like Si or Ge) with $U_{cov}$ being the bonding potential, while the right end represents (almost) purely 'ionic' systems with $U_{ion}$ being the electrostatic bonding potential with some $U_{cov} \ll U_{ion}$. (*ii*) and (*iii*) represent a partially 'covalent' system that was perturbed by a mechanical displacement or an electric field, respectively. $\Delta \tilde{P}_{elect}$ and $\Delta \tilde{P}_{ion}$ represent the electrostatic work ($\Delta U_{ion}$) that is done upon displacement of the electrons or the ions, respectively. $\Delta U_{cov}$ represents the additional work done against the covalent bonding as a result of atomic displacement. The arrows represent the restoring force-vectors due to the above-mentioned perturbations. Upon mechanical displacement, $\Delta \tilde{P}_{ion}$ and $\Delta U_{cov}$ are assumed to be the dominant restoration factors. Upon an electric field: at optical frequencies ($\omega \to \infty$) atoms are assumed to be static (cf. Born-Oppenheim approximation), so only the electronic hard-sphere is displaced; at low frequencies ($\omega \to 0$), where atoms are allowed to rearrange, both the ionic and electronic hard spheres can be displaced.

To a (limited) extent, the bandgap 'deformation potential':

Eq. 10) $\qquad D_p \equiv \dfrac{\Delta E_g}{\left(\frac{\Delta V}{V}\right)}$

which is the relative change between the VBM and the conduction band minimum (CBM) (or the change in the bandgap, $\Delta E_g$) for a given strain may be a measurable quantity for defining the type of covalence. For that we must assume that:

(a) although the bandgap in a semiconductor can have both a covalent and ionic component - $E_g{}^2 = \left(E_{g_{cov}}{}^2 + E_{g_{ion}}{}^2\right)$,[2] if we assume that upon displacement the degree of ionicity change is (energetically) much smaller than the covalent counterpart (i.e., $\Delta E_{g_{cov}} \gg \Delta E_{g_{ion}}$) [(see Endnote viii)], and

(b) the relative change of the VBM energy will be quantitatively very different from that of the CBM, meaning: $\Delta E_{VBM} \gg \Delta E_{CBM}$ or $\Delta E_{VBM} \ll \Delta E_{CBM}$.

With these assumptions we can correlate:

Eq. 11)    $\Delta \tilde{U}_{cov} \propto D_p$.

Since $\frac{\Delta V}{V}$ is related to the change in the applied pressure change, $\Delta p$, via the bulk modulus, $B$, as: [(see Endnote ix)]

Eq. 12)    $\Delta p = -B \cdot \frac{\Delta V}{V}$

Therefore, $D_p$ is often written as:[17]

Eq. 13)    $D_p = -B \cdot \frac{\Delta(E_g)}{\Delta p}$

It is common to think that $D_p$ is usually negative, which is a result of 'bonding' VBM and an 'anti-bonding' CBM, [(see Endnote x)] and it is indeed the case for materials with diamond-, zincblende- or wurtzite-like structures.[17,18] All these systems are tetrahedrally-coordinated (or, as will be mentioned from now on, their coordination number (CN) is CN=4). However, systems with CN>4, like rocksalt (CN=6), CsCl-like (CN=8) or perovskite (CN=6,12), tend to have a completely different set of orbital hybridization, where the VBM has, in many cases, an 'anti-bonding' character.[5,19–25] The apparent outcome of this is that (to some extent) the algebraic sign of $D_p$, or, as usually measured, $\frac{\Delta(E_g)}{\Delta p}$, corresponds to the *type* of covalent bonding, as summarized in Figure 2 (and further elaborated in the ESI - Section A).

Rewriting Eq. 9 and using Eq. 6, Eq. 8, Eq. 13, we can write the following relation:

Eq. 14)    $\Delta \tilde{U}_{cov} = \Delta \tilde{U}_{disp} - \Delta \tilde{U}_{ion} \approx$

$$\approx \left( B \cdot \Delta V - \varepsilon_{(\omega)} \cdot E_{(\omega)}^2 \cdot V \right) \approx -B \cdot \frac{\Delta(E_g)}{\Delta p}$$

---

viii   Meaning that the change in the overlap between orbitals upon displacement is what mainly influences $E_g$

ix   Since the interatomic spacing can change also with temperature (e.g., cooling results in compression), we can find $D_p$ also via the thermal expansion coefficient, $\alpha_T \equiv \frac{d(\ln(V))}{dT}$, as: $D_p = \frac{1}{\alpha_T} \cdot \frac{d(E_g)}{dT}$. Due to additional thermal effects, $D_p\left(\frac{1}{\alpha_T}\right)$ may be different from $D_p(B)$.

x   Simplistically it can be imagined as $\sigma(sp^3\text{-}sp^3)$ and $\sigma^\star$ orbitals, which is correct for diamond-, zincblende- or wurtzite like systems. A more accurate picture can be found in refs. [17,18].

By dividing with $B$ and then using Eq. 12, we get:

Eq. 15)
$$\left(1 - \boldsymbol{\varepsilon}_{(\boldsymbol{\omega})} \cdot \left(\frac{E_{(\omega)}^2}{B}\right) \cdot \frac{V}{\Delta V} - 1\right) \cdot \Delta V =$$

$$= \left(1 - 1 - \boldsymbol{\varepsilon}_{(\boldsymbol{\omega})} \cdot \left(\frac{E_{(\omega)}^2}{\Delta p}\right)\right) \cdot (-\Delta V) \approx \frac{\Delta(\boldsymbol{E_g})}{\Delta \boldsymbol{p}}$$

In Eq. 15 we have three intensive *measurable* parameters: $B$, $\varepsilon_{(\omega)}$ and $\frac{\Delta(E_g)}{\Delta p}$. Assuming compressive displacement, where $\Delta V < 0$ and $\Delta p > 0$, the only parameter that can change sign is $\frac{\Delta(E_g)}{\Delta p}$, while the others are always positive. $\left(\frac{E_{(\omega)}^2}{\Delta p}\right)$ can be understood as the strength of electric field applied with respect to the material's stiffness. With increasing softness ionicity of a materials, we can expect that $\left(\frac{E_{(\omega)}^2}{\Delta p}\right)$ will grow to large values, which means that under compressive displacement, $\frac{\Delta(E_g)}{\Delta p}$ is expected to get lower (and possibly negative) values. [see Endnote xi]

Therefore, by setting $\Delta V$ as constant:

- The *smaller* $\left(\frac{E_{(\omega)}^2}{\Delta p}\right)$ (i.e., stiffer and/or more covalent) and/or $\varepsilon_{(\omega)}$ (less 'polarizable'), the more *positive* $\frac{\Delta(E_g)}{\Delta p}$ becomes. A positive $\frac{\Delta(E_g)}{\Delta p}$ will correspond to *attractive* ('bonding') covalent interactions in the valence band.

- The *larger* $\left(\frac{E_{(\omega)}^2}{\Delta p}\right)$ (i.e., softer and/or more ionic) and/or $\varepsilon_{(\omega)}$ (more 'polarizable'), the more *negative* $\frac{\Delta(E_g)}{\Delta p}$ becomes. A negative $\frac{\Delta(E_g)}{\Delta p}$ will correspond to *repulsive* ('anti-bonding') covalent interactions in the valence band.

---

xi The parameter $\left(\frac{E_{(\omega)}^2}{\Delta p}\frac{E_{(\omega)}^2}{B}\right)$ can be related to the electrostriction coefficient, $M_{ii}$, since in isotropic materials, the longitudinal electrostrictive strain, $S$ is:[26] $S_{ii} = M_{ii}E^2$. This equation can be rewritten as: $\frac{E^2}{S_{ii}} \cdot \left(\frac{E_Y}{E_Y}\right) = \frac{E^2}{\sigma_L} \cdot (E_Y) = \frac{1}{M_{ii}}$, where $E_{Young's}$ is the Young's elastic modulus and $\sigma_L$ is the longitudinal applied stress. Using the relation between strain and stress and the relation of $B$ to the longitudinal elastic modulus (Young's modulus) and the Poisson ratio to relate linear to bulk mechanical transformations,) via Poisson's ratio ($B = \frac{E_{Young's}}{3-2\cdot v}$),[27] Eq. 15 can be become of the form written as $\left(1 - \boldsymbol{\varepsilon}_{(\boldsymbol{\omega})} \cdot \left(\frac{f(v)(3-2\cdot v)\cdot \Delta p}{E_Y B^2 \cdot M_{ii}}\right)\right) \cdot |\Delta V|(-\Delta V) \approx \frac{\Delta(E_g)}{\Delta p}$, where $f(v)$ is a Poisson-related function. This new relation suggests that a large $M_{ii}$ may be another indication for 'anti-bonding' repulsive' nature of the interatomic bonds. We note this possible correlation as a trigger for future study. We also note that these relations should be used as qualitative guidelines rather than exact analytical relations.

Therefore, we see that *repulsive* covalent character is more likely in 'soft', highly ionic (where CN > 4) [see Endnote xii] systems, but with rather large $\varepsilon_{(\omega)}$. Since we consider here processes where ionic relaxation is allowed, $\varepsilon_{(\omega)}$ can be referred to $\varepsilon_{(\omega \to 0)} \equiv \varepsilon_s$.

Following Figure 2(i) (and **Fig. S1**), systems with $D_p > 0$ (or $\frac{\Delta(E_g)}{\Delta p} < 0$) are usually reported as having 'anti-bonding' VBM orbitals.[5,19–25] As will be presented later in Figure 3 in the Results section – these also usually possess relatively ionic character, with a significant energetic overlap between neighboring atoms that allows orbital mixing as reported elsewhere.[5,19–25] Following **Fig. S1**(ii), 'classical' semiconductors with *attractive* covalent character appear to be with CN=4, while those with *repulsive* covalent character are (in most cases coordinated with a higher number, i.e., CN>4. [see Endnote xiii]

Now let us consider a variation to Eq. 9 using Eq. 8 and Eq. 13 :

Eq. 16)  $\qquad \Delta \breve{P}_{(\omega)} = \Delta \widetilde{U}_{ion} = \Delta \widetilde{U}_{disp} - \Delta \widetilde{U}_{cov} \approx B \cdot \left( \Delta V + \frac{\Delta(E_g)}{\Delta p} \right)$

Considering a constant polarization energy, $\Delta \breve{P}_{(\omega)}$, we learn from Eq. 16 that when $\frac{\Delta(E_g)}{\Delta p}$ is more negative (i.e., less covalent bonding), the allowed atomic displacement, $\Delta V$, increases. Considering the frequency dependence, following Figure 1 and Eq. 16(iii), we see that at optical frequencies ($\omega \to \infty$) the ions are practically static (cf. Born–Oppenheimer approximation) and only the electronic hard-sphere can be displaced, leading to '*Electronic polarizability*' (EP). Thus, because $\Delta \breve{P}_\infty$ is the only contribution to $\Delta \widetilde{U}_{ion}$, using Eq. 6:

Eq. 17)  $\qquad \Delta \breve{P}_\infty = \Delta \breve{P}_{elect} \propto \varepsilon_{elect} \approx \varepsilon_\infty$

However, at low frequencies ($\omega \to 0$), where nuclei have sufficient time to respond to an electric field and be displaced from their equilibrium position, '*structural polarizability*' (SP) also contributes to $\Delta \widetilde{U}_{ion}$, so that:

Eq. 18)  $\qquad \Delta \breve{P}_s = \Delta \breve{P}_{elect} + \Delta \breve{P}_{ion} \propto \varepsilon_{elect} + \varepsilon_{ion} \approx \varepsilon_s$

$\varepsilon_\infty$ and $\varepsilon_s$ are two often-measured parameters. $\varepsilon_\infty$ can be measured from, for example, the optical refractive index, while $\varepsilon_s$ is often measured using impedance spectroscopy at DC to ~kHz frequencies. [see Endnote xiv]

---

xii   When $U_{cov} > U_{ion}$, the cohesive energy is mostly dominated by covalent bonds, $U_{cov}$, and should be strongly attractive ('bonding'), otherwise the structure should not be stable. When, $U_{cov} < U_{ion}$, the cohesive energy is mostly electrostatic so $U_{cov}$ can also have repulsive character. Related to this point, if ionicity increases above a certain value ($f_i^{Phillips} > 0.785$), the structural symmetry changes to a higher coordination number (zincblende/wurtzite to rocksalt. The reason is that cohesive energy increases with CN when bonding is mostly electrostatic (i.e., large $U_{ion}$).[2]

xiii  In textbooks, 'classical' covalent semiconductors are usually structured as a tetrahedrally-coordinated (CN=4) network of *hybridized sp³-sp³* 'bonding' orbitals that form a closed-packed FCC (e.g., zincblende- or wurtzite-like) structure. With decreasing covalence towards a more ionic system the bandgap usually increases, making 'semiconductors' more 'insulators' as well as changing their CN to > 4.

xiv   We consider only elastic processes and not plastic processes such as ion migration.

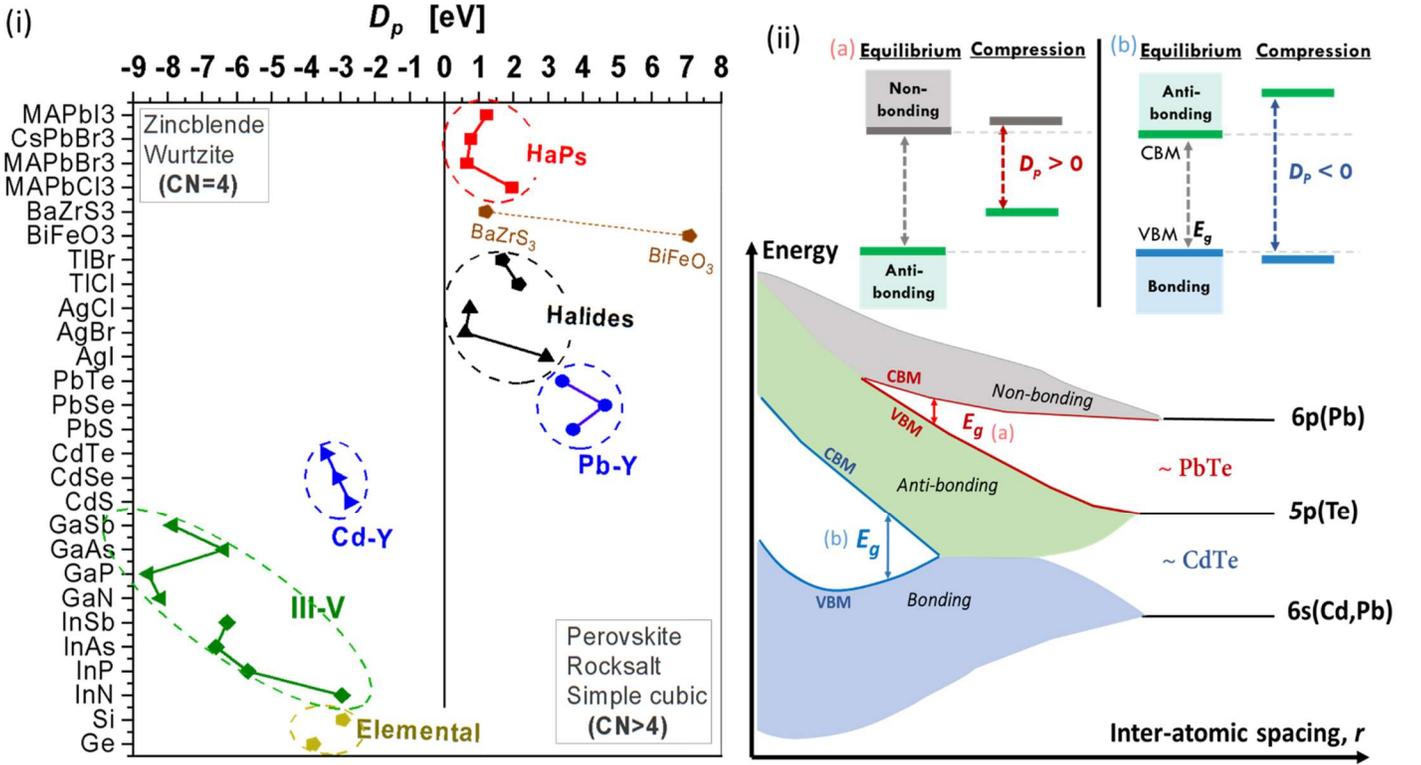

Figure 2: (*i*) Deformation potentials, $D_p$, of different crystalline frameworks. $D_p$ values are derived using Eq. 13 and the experimentally-derived values are presented in Table_S 1, together with the relevant references. The shown values are for those reported at room-temperature. The more frequently reported, $\frac{\Delta(E_g)}{\Delta p}$, which is proportional to $(-D_p)$ is plotted in **Fig. S1**(i). Among the ~25 heteropolar and ~2 homopolar chosen materials, all of the materials with CN=4 are with $D_p < 0$ (and $\frac{\Delta(E_g)}{\Delta p} > 0$), while all of those with CN>4 (except wurtzite $\beta$-AgI with CN=4) are with $D_p > 0$ (and $\frac{\Delta(E_g)}{\Delta p} < 0$) (see **Fig. S1**(ii)). (*ii*) Schematic representation of the development of the band structures as a function of the inter-atomic spacing, usually derived from tight-binding models (see for example ref. [28] chapter 4). We chose CdTe and PbTe as two systems with identical valence of both cations and anions, but with a very different structure and bonding type. These systems illustrate the development of orbital hybridization in 'classical' $sp^3$-like systems (e.g., Si ,GaAs, CdTe – see ref. [17] for example), and that of an 'anti-bonding' VBM (like HaPs or Pb-chalcogenides)[5,19–25]. The r value for the vertical double arrows (representing $E_g$) were chosen, based on the relative interatomic spacing ($r_{(Cd-Te)} < r_{(Pb-Te)}$), and are for illustrative purposes only. As further illustrated in sub-figures (ii, a) and (ii, b), $D_p$ is expected to be *positive* for the 'anti-bonding' VBM and *negative* for the 'bonding' VBM case, respectively.

Following Eq. 16, to estimate the covalence *type* and *degree*, we have to allow atomic relaxation, meaning that we need to gain insight into $\Delta \breve{P}_{ion}$. To do so, we use the SP to EP ratio to understand the relative contribution of SP to the total polarization, and use the term '*relative structural polarizability*' (RSP). Using Eq. 17 and Eq. 18, RSP can be represented as:

Eq. 19) $\qquad RSP \propto \frac{\varepsilon_{ion}}{\varepsilon_{elect}} = \left(\frac{\varepsilon_{ion} + \varepsilon_{elect}}{\varepsilon_{elect}} - 1\right) = \left(\frac{\varepsilon_s}{\varepsilon_\infty} - 1\right)$

To get a better understanding of how the RSP can be indicative for the degree and type of interatomic bonding, we consider (purely) covalent and (mostly) ionic systems, meaning $\Delta U_{cov} \rightarrow 1$ and $\Delta U_{cov} \rightarrow 0$, respectively. In these cases, one should expect $\left|\frac{\Delta(E_g)}{\Delta p}\right| \gg 0$ and $\frac{\Delta(E_g)}{\Delta p} \rightarrow 0$, respectively. In the purely covalent case, there will be no depolarization contribution from the ions (i.e., $\Delta \breve{P}_{ion} \rightarrow 0$), so one should expect that $\Delta \breve{P}_\infty \approx \Delta \breve{P}_s$, and observe $\varepsilon_s = \varepsilon_\infty$, or $\left(\frac{\varepsilon_s}{\varepsilon_\infty}-1\right)$=0. This is indeed the case for covalent materials like Si or Ge (see Table_S 3: Parameters, used to extract the 'covalence ' using Pauling's 'resonating bond' model2, following Eq. 2. *M* and *N* are for the room-temperature phases. XA and XB are taken from three different sources (see top row for references) and plotted against RSP $\frac{\varepsilon_s}{\varepsilon_\infty}-1$ in Fig. S2. XA and XB for perovskite structures were related to the B cation and the X anion, respectively, since these are the atoms closest to each other in perovskite structures).

In the case of (mostly) ionic systems, where $\widetilde{U}_{ion} \gg \widetilde{U}_{cov}$, electrostatic interactions will govern the energy balance. Because additional (attractive or repulsive) covalent interactions can be neglected at $\omega \rightarrow 0$, there should be no energetic difference between displacement of the electronic hard-shell of the ions at low or high frequencies. This suggests that $\Delta \breve{P}_{elect} \rightarrow \Delta \breve{P}_{ion}$, and consequently, $\varepsilon_{ion} \rightarrow \varepsilon_{elect}$, where following Eq. 18, this means that $\varepsilon_s \rightarrow 2\varepsilon_\infty$ or $\left(\frac{\varepsilon_s}{\varepsilon_\infty}-1\right) \rightarrow 1$. The experimental values of $\varepsilon_s$ and $\varepsilon_\infty$ for systems that are classically referred to as 'ionic' (like KBr, CsI or ZnO (see Table_S 3: Parameters, used to extract the 'covalence ' using Pauling's 'resonating bond' model2, following Eq. 2. *M* and *N* are for the room-temperature phases. XA and XB are taken from three different sources (see top row for references) and plotted against RSP $\frac{\varepsilon_s}{\varepsilon_\infty}-1$ in Fig. S2. XA and XB for perovskite structures were related to the B cation and the X anion, respectively, since these are the atoms closest to each other in perovskite structures)), show that for these indeed $\varepsilon_s \rightarrow 2\varepsilon_\infty$.

For intermediate cases, where both ionic and covalent characters are important, we suggest (now equipped with experimental justifications for the boundary conditions) that the RSP (Eq. 19) will vary with the *type* and *degree* of the covalent character. If the covalent character is *attractive* (in addition to that of the electrostatic one), we expect $\Delta \breve{P}_{elect} > \Delta \breve{P}_{ion}$, and consequently, $\varepsilon_s < 2\varepsilon_\infty$ or RSP < 1. If the covalent character is *repulsive,* the structural polarizability will dominate over the electronic one (or $\Delta \breve{P}_{ion} > \breve{P}_{elect}$) and, consequently, $\varepsilon_s > 2\varepsilon_\infty$ or RSP > 1.

In summary, we showed that $\frac{\Delta(E_g)}{\Delta p}$ correlates with the *type* of covalence (less with the *degree* of covalence), and changes sign with between materials having with CN=4 and those with CN>4. RSP seems to give both qualitative and quantitative measure to the type and degree of covalence of a system. We summarize in Table I the correlations that should be found for materials with different bond character.

Table I: Summary of the correlations between the coordination number (CN), the bandgap-pressure coefficient (i.e, $\left(\frac{\Delta(E_g)}{\Delta p}\right)$) and the RSP (i.e., $\left(\frac{\varepsilon_s}{\varepsilon_\infty} - 1\right)$), as proposed in the main text. RSP uniquely correlates with the degree and type of covalence.

| Interatomic bonding type: | CN | $\frac{\Delta(E_g)}{\Delta p}$ | RSP $\left[= \left(\frac{\varepsilon_s}{\varepsilon_\infty} - 1\right)\right]$ |
|---|---|---|---|
| with a *repulsive* ('anti-bonding') *covalent* nature | > 4 | < 0 | > 1 |
| *Purely ionic* | > 4 | → 0 | → 0 |
| with an *attractive* ('bonding') covalent nature | = 4 | > 0 | < 1 |
| *Purely covalent* | = 4 | > 0 | → 0 |

### III.  Results:

#### A.  Model verification

To check how well our model works, we first try to match it with classical models for 'covalence'. To that end we compare the RSP with Pauling's classical definition for 'ionicity' or $f_c'$ (1-'ionicity'), following Eq. 2. By doing that, we see in (Figure 3) a clear (empirical) [see Endnote XV] correlation between RSP and $f_c'$, which indicates that RSP follows the classical concepts of 'ionicity'. It is interesting to see that classical ionic materials (e.g., KBr) converge around RSP=1 as predicted. The most interesting part is the correlation with materials with RSP>3, which are known, theoretically, to possess 'anti-bonding' type of covalence at their VBM.[5,19–25] As shown in Figure 2, and further elaborated in Figure 4, these same materials that have RSP>3 also show $\frac{\Delta(E_g)}{\Delta p} < 0$, which indicates a rough qualitative estimate for the type of covalence, as explained earlier.

While RSP and $\frac{\Delta(E_g)}{\Delta p}$ correlate well by threshold of value and sign, respectively (following **Table I**), Figure 4 shows that the quantitative correlation is lost at the 'repulsive' side of covalence (this may result due to the variety of structural geometries with respect to covalently-attractive (CN=4) systems). Moreover, as can be seen in Table_S 1 and Table_S 3: Parameters, used to extract the 'covalence ' using Pauling's 'resonating bond' model2, following Eq. 2. $M$ and $N$ are for the room-temperature phases. XA and XB are taken from three different sources (see top row for references) and plotted against RSP $\frac{\varepsilon_s}{\varepsilon_\infty} - 1$ in Fig. S2. XA and XB for perovskite structures were related to the B cation and the X anion, respectively, since these are the atoms closest to each other in perovskite structures, $\frac{\Delta(E_g)}{\Delta p}$ does not indicate a clear difference between purely covalent materials, like Si or Ge, and covalent III-V semiconductors; RSP, on the other hand, does show a significant difference.

---

The values for $\frac{\Delta(E_g)}{\Delta p}$, $\varepsilon_s$, $\varepsilon_\infty$ and electronegativity $X_i$, which are derived from experimental observables, are presented in Table_S 1 and Table_S 3. When $\frac{\Delta(E_g)}{\Delta p}$ was not explicitly mentioned, we derived $\frac{\Delta(E_g)}{\Delta p}$ from the experimental data following the procedure described in **Fig. S7** (ESI - Section A). While values for $X_i$ are known to vary between different approaches of electronegativity, as can be seen from the tables in Huheey in 1932 based on calorimetry (Pauling), in 1968 based on dielectrics (Phillips[2,30] - much different approach than ours, though) and till 2015 from simulations (Lung & Smith[29]), we find that all correlate similarly-well with RSP (see **Fig. S2**).

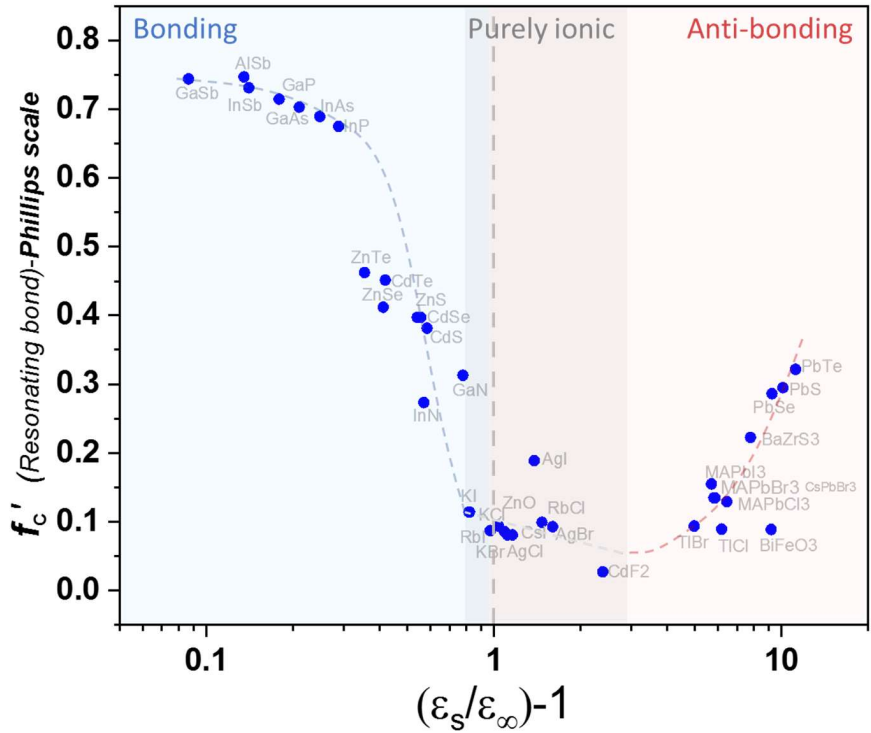

Figure 3: A semi-log correlation between Pauling 'covalence', $f_c'$ (Eq. 2), and RSP, $\left(\frac{\varepsilon_s}{\varepsilon_\infty} - 1\right)$ (Eq. 19). Dielectric constants and electronegativity values are summarized in Table_S 2 and Table_S 3. Electronegativity values are on the corrected Phillips[2,30] scale, which is based on dielectric properties. Similar correlations, but with different values of electronegativity (based on the original Pauling (1932)[1] values and those of Lung & Smith (2015)[29]), are presented in **Fig. S2**. All correlation plots show a similar picture, where the one based on dielectric properties (by Phillips) correlates with RSP better than the others. For perovskite structures, we used the electronegativity values for the B cation and X anion, as in the ABX$_3$ composition this is the shortest bond and, thus, most dominating backbone of the structure. The dashed lines are guides for the eye following the correlations. The light gray area was arbitrarily chosen to be between (0.8-2) as an area where materials are represented as 'purely ionic'.

Following these results that support our model, we suggest RSP as a *fundamental* observable for understanding the essence of interatomic bonding in semiconductors. [(see Endnote xvi)] As far as we know, this is the first time that 'anti-bonding' covalence is quantified using experimental observables. As such, this work adds the type of covalence to the degree of covalence as a parameter that can be estimated from experimentally observable measurables.

The correlation between RSP and $\frac{\Delta E_g}{\Delta p}$ (Figure 4) is less pronounced, but with a clear qualitative picture that all cases with $\frac{\Delta E_g}{\Delta p} > 0$ have RSP < 1, while those with $\frac{\Delta E_g}{\Delta p} < 0$ always show RSP > 1. At the limit where RSP → 1, where covalence becomes negligible and the system would be 'purely-ionic', $\frac{\Delta E_g}{\Delta p}$ indeed goes to zero. This correlation suggests that, similar to RSP, $\frac{\Delta E_g}{\Delta p}$ *indeed* reflects the type of the covalent bonding (though, with much less qualitative accuracy, if any). Several theoretical works, that aimed to show chemical trends to estimate the degree of the covalence,[17,18] were limited to tetrahedrally-bonded systems, except for ref.[19], which dealt with internal trends among Pb-chalcogenides. To the best of our knowledge, analysis or discussions about chemical trends for $D_p$ or $\frac{\Delta E_g}{\Delta p}$ of other 'anti-bonding' sets of materials (other than the above-mentioned Pb chalcogenides) are absent.

It is interesting to note the relation between the type of covalence and the structural symmetry (or the CN), which was found to correlate with $\frac{\Delta E_g}{\Delta p}$ (see **Fig. S1**(ii)). Based on most of our data set, we see that systems with CN>4, tend to have anti-symmetric valence wavefunctions, which arise from the interaction between 'atomic' orbitals (e.g., s--p). This differs from interactions between 'molecular' orbitals (e.g., sp³--sp³), which happen in CN=4 systems. An exception, however, is $\beta$-AgI, which has a wurtzite symmetry, but with 'anti-bonding' covalence, the origin of which warrants a separate study.

Usually, materials with high ionicity will tend to form structures with higher CN, as it increases their cohesive energy[2] due to an increase of charge delocalization - (cf. also comments *xii* and *xiii*). We postulate that existence of 'anti-bonding' valence orbitals can exist in materials where the electrostatic (Coulomb) attraction is sufficiently large, so it suffices to energetically favor the existence of compounds having repulsive covalent character. This condition appears to be fulfilled in materials with CN>4. With a similar logic, in systems in which covalence dominates, such as tetrahedrally-coordinated sp³-sp³ bonded systems, covalence is allowed to be only an attractive one.

---

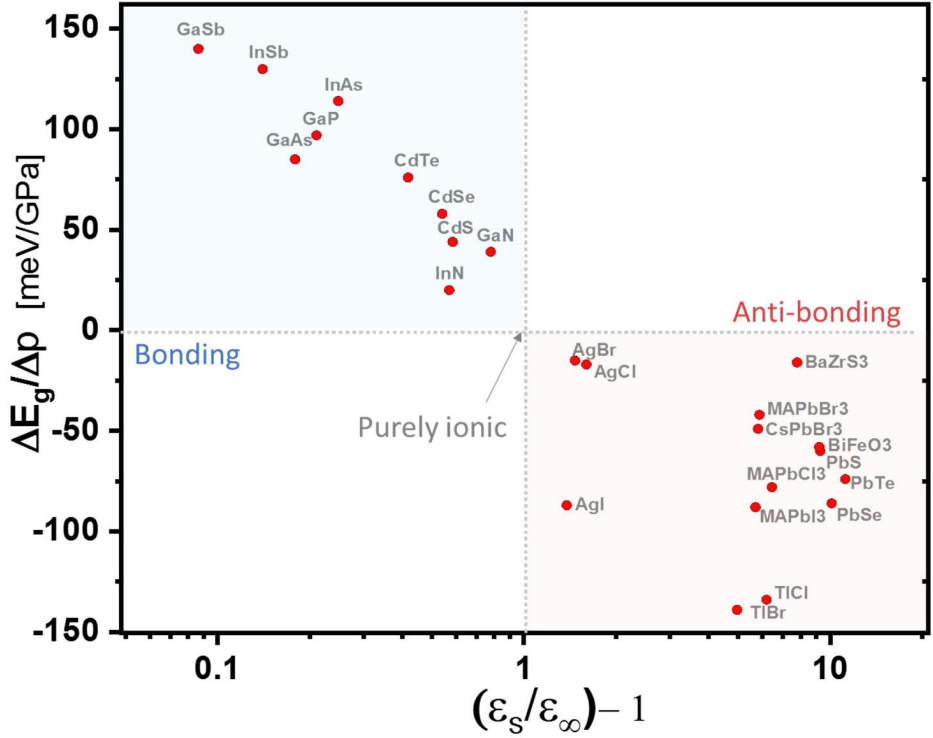

Figure 4: A semi-log correlation between the bandgap-pressure coefficient $\frac{\Delta(E_g)}{\Delta p}$ and RSP $\left(= \frac{\varepsilon_s}{\varepsilon_\infty} - 1\right)$. The plot is consistent (at least qualitatively) with the postulates of Table I. When the VBM has a *repulsive* ('anti-bonding') covalent character, $\left(\frac{\varepsilon_s}{\varepsilon_\infty} - 1\right) > 1$ (or $\varepsilon_s > 2 \cdot \varepsilon_\infty$) and $\frac{\Delta(E_g)}{\Delta p} < 0$. When the VBM has an *attractive* ('bonding') covalent character, $\left(\frac{\varepsilon_s}{\varepsilon_\infty} - 1\right) > 1$ (or $\varepsilon_s < 2 \cdot \varepsilon_\infty$) and $\frac{\Delta(E_g)}{\Delta p} > 0$. The transition point from 'bonding' to 'anti-bonding' covalently bonded materials passes through the point where materials are 'purely ionic' with $\left(\frac{\varepsilon_s}{\varepsilon_\infty} - 1\right) = 1$ (or $\varepsilon_s = 2 \cdot \varepsilon_\infty$) and, following the trend from the 'bonding' side (but less obviously from the 'anti-bonding' side) $\frac{\Delta(E_g)}{\Delta p}$=0. Purely covalent cases of Si and Ge, where $\left(\frac{\varepsilon_s}{\varepsilon_\infty} - 1\right) = 0$ (or $\varepsilon_s = \varepsilon_\infty$), are $\frac{\Delta(E_g)}{\Delta p} > 0$ is fully consistent with Table I. Following **Fig. S1**(ii), since most materials with CN=4 show $\frac{\Delta(E_g)}{\Delta p} > 0$ and all CN>4 show $\frac{\Delta(E_g)}{\Delta p} < 0$, we conclude that CN=4 materials relate to attractive-covalence RSP, while CN>4 relates to strongly ionic and repulsive-covalence RSP. The points for Si and Ge are not presented since the x-axis is on a log scale. The values (all based on experimental data) related references are presented in Table_S 1 and Table_S 2.

Repulsive covalent interactions must induce 'geometric frustration', meaning that 'anti-bonding' covalence comprises two (energetically) competing processes: (1) maintenance of high structural symmetry (to gain entropy) and (2) breaking this symmetry due to the repulsive nature of the bonds. Good examples for this competition are the 1st and 2nd Jahn-Teller (JT) effects.[34] Upon *static* distortion, where broken symmetry is permanent over time, the gain in electronic energy due to a broken symmetry dominates over the gain in entropy and 1st order JT distortion is an example of such phenomena. In a *dynamic* picture, however, the system is locally and over a short period of time (in the order of few natural structural vibration periods) distorted, but over long acquisition times, due to gain in entropy, its symmetry remains high, which is an example of $2^{nd}$ order JT distortion. If we consider the analogy between 'anti-bonding' covalent bonding and dynamically distorted systems, we see a clear similarity, as in both cases there are repulsive forces that will drive the system to distort from its equilibrium and break the symmetry.

Indeed, systems like HaPs are known to show features of dynamic disorder.[35,36] Similar to HaPs, which are corner-sharing polyhedral structures with a high probability for anharmonic motions, lone-pair effects in Pb-chalcogenides[37] and Tl-halides[20] are also known to induce dynamic disorder. We conclude that compounds that are: (1) highly coordinated (CN> 4), (2) connected via corner-sharing polyhedra, (3) contain heavy elements and/or (4) possess a sterically distorting lone-pair, show 'anti-bonding' type of behavior with RSP > 1 (and $\frac{\Delta E_g}{\Delta p} < 0$). From this viewpoint, whatever the reason leading to dynamic structural distortion, we can consider it as 'repulsive' covalence.

A somewhat different, but related effect is the static dielectric response in paraelectric systems (e.g., $SrTiO_3$)[38] or during depolarization of spontaneously-polarized ferroelectric materials. Considering the real part of the dielectric function, [see Endnote xvii] the overall contribution to $\varepsilon$ in heteropolar systems is known to result from three (main) types of charge displacements: *electronic* (due to motion of the electronic cloud), *dipole reorientation* (i.e., relative reorientation of an ionic pair with respect to an electric field) and *ionic* (due to the relative motion between oppositely-charged atoms). In this study, however, we neglect 'dipole reorientation', which is justified in isotropic solids. [see Endnote xviii] Whenever dynamic dipoles exist,[40] $\varepsilon_s$ is usually found to be large (on the order of thousands). [see Endnote xix] In ferroelectric materials,[40] where during depolarization (above coercive and below saturation electric fields), rearrangement of dipoles leads to high dielectric responses. [see Endnote xx]

Common to all these cases is that a system has a degenerate set of energies, where 'geometric' or 'electronic' frustration leads to structural distortions. Therefore, when enthalpy overcomes entropy, the system

---

xvii  We neglect inelastic processes, which will contribute to the imaginary part of the dielectric function.

xviii  Contributions due to dipole reorientation may become important in polar crystals and may depend on the direction of the applied electric field (as found for GaN)[39].

xix  These high values of $\varepsilon_s$ show dipole-dipole interactions, where long-range dipole interactions lead to a collective contribution to the dielectric response.[41] As temperature increases, de-coherence (due to phonons) leads to a decrease in $\varepsilon_s$. This is demonstrated for $SrTiO_3$, which is known to be paraelectric, where with temperature increase, the dielectric constant drops[38] while lattice distortion increases.[42] Relaxor-ferroelectric materials are similar, but their coherence is dynamic and much stronger depend on temperature (in addition to frequency).[41] The temperature-dependence relates to an inherent anharmonicity of these systems.

xx  Apart from ferro- and para-electric materials or materials with JT effect, materials that show 'ligand hole' effects are also known to be dynamically disordered.[43,44] The common ground for all these cases is that a system has a degenerate set of energies, where 'geometric' or 'electronic' frustration leads to structural distortions. When enthalpy overcomes entropy, 'frozen states' start to dominate, and ferroelectric phases or first-order Jan-Teller effects appear.

will be static and one may observe ferroelectricity or a $1^{st}$-order JT effect. If entropy overcomes enthalpy, the system will become 'dynamic in time' and it is more reasonable to find paraelectric and $2^{nd}$-order JT effects.

Paraelectric systems (like $SrTiO_3$), where collective dipole interactions give rise to huge $\varepsilon_s$, and show RSP>>1. This case, however, has an origin that is different than the repulsive interactions due to anti-bonding orbital or effects to degeneracy-breaking phenomena, such as a $2^{nd}$ order JT effect. The effect of the repulsive covalence on the RSP is, however, much smaller than collective dielectric responses and should increase with increasing temperature, while the RSP of paraelectric systems should show a decrease with increasing temperature.

### B. Implications: How does the covalent nature affect electronic effective mass, perturbation screening and mobility?

In semiconductors, free charges will flow in a band (electrons in the CB and holes in the VB) with a specific effective mass and will scatter due to perturbations of the potential landscape they travel in. The '*electronic effective mass*', $m^*$, and '*mobility*', $\mu$, for free charges are often referred to as fundamental figures of merits of free charge dynamics in semiconductors. With values for RSP in hand, we can start to see the importance of the describing bonding in terms of both type and degree of covalence, on charge dynamics.

The effective mass, $m^*$ depends mostly on the orbital coupling in a compound, which implies that it should also depend on its covalence (and thus, its RSP). Generally, $m^*$ is derived from the energy-momentum relation close to the band extrema, or more specifically, the curvature of the bands at the VBM and CBM ($m^* \sim \frac{\partial^2 E}{\partial k^2}$, where $k$ is momentum). Stronger interatomic orbital mixing usually results in smaller $m^*$.[45] Using RSP as a parameter which indicates the degree of covalence, and collecting from the literature experimentally-derived effective mass values (usually from cyclotron resonance or magneto-resistance experiments – see values in Table_S 4), we show in Figure 5 a fairly good correlation with RSP (and, to some limited extent, also with $\frac{\Delta E_g}{\Delta p}$ – see **Fig. S4**). Figure 5 shows that $m^*$ decreases with the degree of covalence – *regardless of the type of covalence*. Moreover, as expected from our model, it is reassuring to find that materials that are mostly ionic show a maximum $m^*$ value, when RSP→1 (or $\frac{\Delta E_g}{\Delta p}$ →0 in **Fig. S4**). *We conclude that an increase of covalence (regardless of the type) should decrease a material's $m^*$, and lead to a material with improved electron and hole mobilities – a property that can be reasonably estimated by RSP (and somewhat by $\frac{\Delta E_g}{\Delta p}$).*

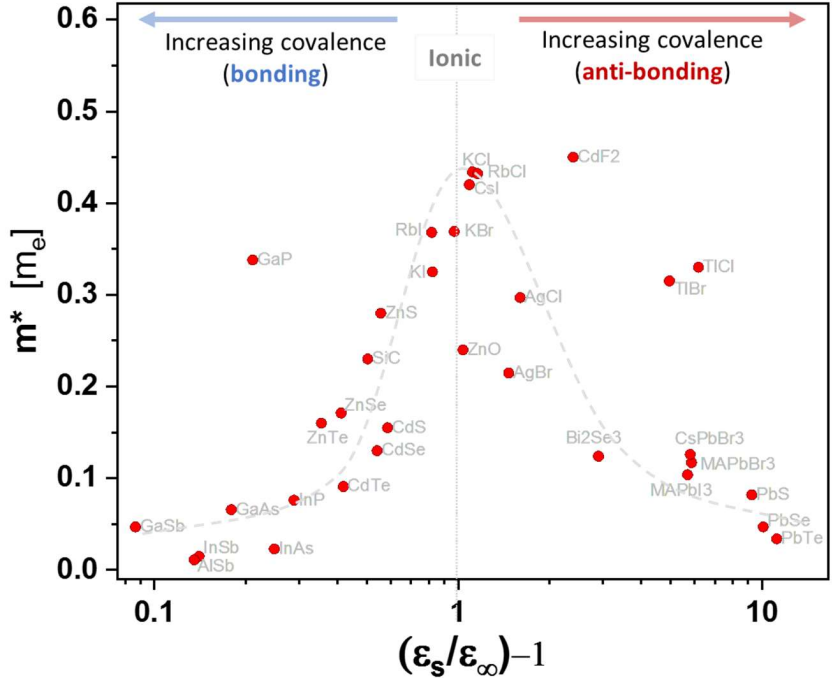

**Figure 5**: Semi-log plot of the electronic effective mass, $m^*$, as function of RSP $\left(\frac{\varepsilon_s}{\varepsilon_\infty} - \mathbf{1}\right)$. The experimental values used are given also in Table_S 2 and Table_S 4. Fig. S4 gives a plot of $m^*$ vs. $\frac{dE_g}{dP}$.

Unlike $m^*$, $\mu$ depends on the natural lattice vibrations (which *are* a property of the bond), and also on structural defects. In heteropolar systems, phonons and point defects are the most common perturbation mechanisms that limit the charge mobility.[4,45] The most common scattering mechanisms are electric fields that are, at higher temperatures, momentarily induced by polar lattice vibrations, and at lower temperatures by charged point defects.[4,45] Therefore, for screening electric fields, we distinguish between pure electronic polarizability (EP), that refers to $\Delta \breve{P}_{elect} \propto \varepsilon_\infty$ and a combination between electronic and structural polarizability, i.e., $\frac{\Delta \breve{P}_{ion}}{\Delta \breve{P}_{elect}} \propto \left(\frac{\varepsilon_s}{\varepsilon_\infty} - 1\right) = RSP$ (see Eq. 17 - Eq. 19). Following our model, and to illustrate the polarizability difference between covalent 'bonding' and 'anti-bonding' materials, we show the correlation between $\frac{\Delta E_g}{\Delta p}$ and either RSP or $\varepsilon_\infty$ (Figure 6), where we refer to $\frac{\Delta E_g}{\Delta p}$ as an indicator of covalence type.

We see that covalent 'bonding' compounds (with $\frac{\Delta E_g}{\Delta p} > 0$) are much more electronically polarizable (Pb-chalcogenides present an exception; see **Fig. S3**), while in 'anti-bonding' compounds (with $\frac{\Delta E_g}{\Delta p} < 0$) structural polarizability is much more dominant. This strongly suggests that, when charges interact with the lattice (scattering or screening), covalent 'bonding' compounds will respond differently in different frequency ranges than 'anti-bonding' ones: screening by ions will be dominant in covalent 'anti-bonding' compounds, while screening by electrons will be dominant for 'bonding' ones. Therefore, for high (optical) frequency applications, covalent 'bonding' materials will tend to be more polarizable, while for timescales of the order of several vibrational cycles (~100's of GHz and shorter), 'anti-bonding' covalent compounds may be more relevant.

After understanding that covalence relates to both the effective mass and the (electronic or structural) polarizability, we can quantify how type and degree of covalence affect charge mobility.

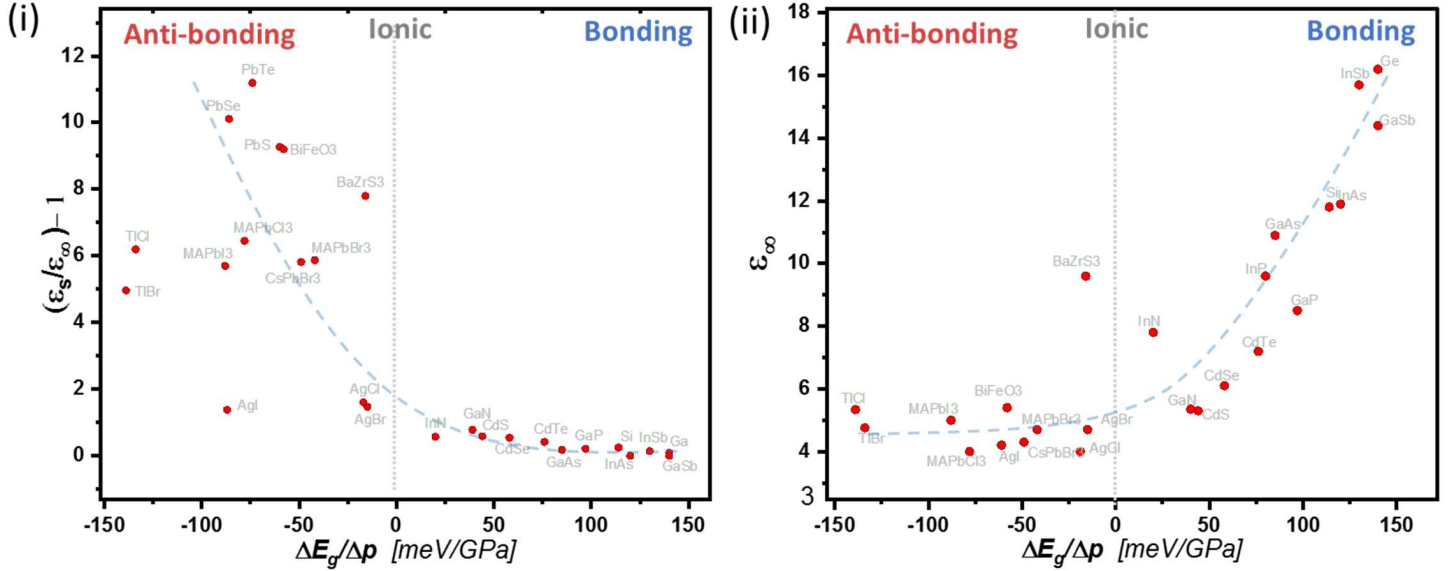

Figure 6: A lin.-lin. correlation between the bandgap-pressure coefficient $\frac{\Delta E_g}{\Delta p}$ and (*i*) the RSP - $\frac{\varepsilon_s}{\varepsilon_\infty} - 1$) and (*ii*) the *electronic* polarizability (EP - $\varepsilon_\infty$). It is clearly seen that the EP is pronounced mostly in covalent '*bonding*' materials, while the RSP is pronounced mostly in covalent 'anti-bonding' materials. Exception compounds (as shown in **Fig. S3**) are Pb-chalcogenides with a very high $\varepsilon_\infty$ (and thus EP). See Table_S 1 and Table_S 2 for numerical values.

### C.  *Mobility* ($\mu$):

Mobility of free charge carriers, in our context, refers to the efficiency of free carriers to migrate in a periodic crystalline lattice; $\mu$ is inversely proportional to the scattering rate of these carriers. The total scattering rate, $\frac{1}{\tau_{total}}$, is a function of all the scattering events: $\frac{1}{\tau_{total}} = \frac{1}{\tau_{S1}} + \frac{1}{\tau_{S2}} + \cdots$ , where $\tau_s$ is the average lifetime of charge carriers between two scattering events. Free charge carrier mobility can then be defined as $\mu_i = \frac{q \cdot \tau_i}{m^*}$, where $q$ is the electronic charge and $m^*$ is the previously presented electronic effective mass. The overall mobility will then be:

Eq. 20)  $\qquad \frac{1}{\mu_{total}} = \frac{1}{\mu_{II}} + \frac{1}{\mu_{NI}} + \frac{1}{\mu_{POP}} + \frac{1}{\mu_{PZ}} + \frac{1}{\mu_{ADP}} + \frac{1}{\mu_{ODP}} + \cdots$

where the different types of $\mu_i$ represent scattering by: **I**onized **I**mpurities (*II*), **N**eutral **I**mpurities (*NI*), phonon-generated **D**eformation **P**otential scattering (*ADP* and *ODP* are for **A**coustic and **O**ptical phonons, respectively), phonon-generated electrostatic potential scattering (*PAP* and *POP* for **P**olar **A**coustic and **O**ptical **P**honons, respectively) and others. The scattering potential, which varies with the specific scattering mechanism, is used to derive $\tau_s$, as can be found in reference [45] (chapters 1 and 2) or reference [4] (chapter 8) for the different scattering mechanisms. We will treat *II*, *NI*, and *POP* scattering potentials as usually the most important scattering mechanisms in heteropolar materials (see refs.[46,47] for specific examples). (see Endnote xxi)

Mobility expressions for *II*, *NI*, and *POP* are commonly expressed as following: [4]

Eq. 21) $\quad\quad \mu_{II} \propto \dfrac{(\varepsilon_s)^2}{m^{*\,1/2}} \cdot \dfrac{T^{3/2}}{N_I}$

Eq. 22) $\quad\quad \mu_{NI} \propto \dfrac{m^*}{\varepsilon_s} \cdot \dfrac{1}{N_N}$

Eq. 23) $\quad\quad \mu_{POP} \propto \dfrac{\varepsilon_{eff}}{(m^*)^{\frac{3}{2}}} \cdot \dfrac{T^{\frac{1}{2}}}{\theta_D} \cdot \left( \exp\left(\dfrac{\theta_D}{T}\right) - 1 \right) \quad ; \quad \varepsilon_{eff} = \left( \dfrac{1}{\varepsilon_\infty} - \dfrac{1}{\varepsilon_s} \right)^{-1} = \dfrac{\varepsilon_s}{RSP}$

where $T$ is temperature, $N_I$ and $N_N$ are ionized and neutral defect densities and $\theta_D$ is the Debye temperature. Following the mobility equations presented above, the extensive parameters are the temperature and the defect density, while the intensive parameters (which can be referred as the fundamental material properties – usually around a given temperature), are $m^*$, the dielectric constants $\varepsilon_s$ and $\varepsilon_\infty$, and $\theta_D$. Assuming fixed temperatures and defect densities, from Eq. 21-Eq. 23 we see that $m^*$, $\varepsilon_s$ and RSP (marked in red in Eqs. 21-23) are analytically dominant and can give a good estimate of the overall mobility. Using the parameters we collected, we plot in Figure 7 the parameters marked in red in Eq. 21-Eq. 23 against RSP, our measure for the type and degree of covalence. Amazingly, we find ~*exponential* dependence between the *degree* of covalence and the charge mobility, but not a large qualitative difference between 'bonding' and 'anti-bonding' covalent materials. This leads us to the conclusion that *any* covalent characteristic in a structure should *improve* charge mobility (Figure 7 (i) and (iii)). An exception is given by materials in which neutral defects are dominant (highly (statically) disordered materials as an example), which show an increase in mobility with the increase of the material's ionicity (Figure 7 (ii)).

---

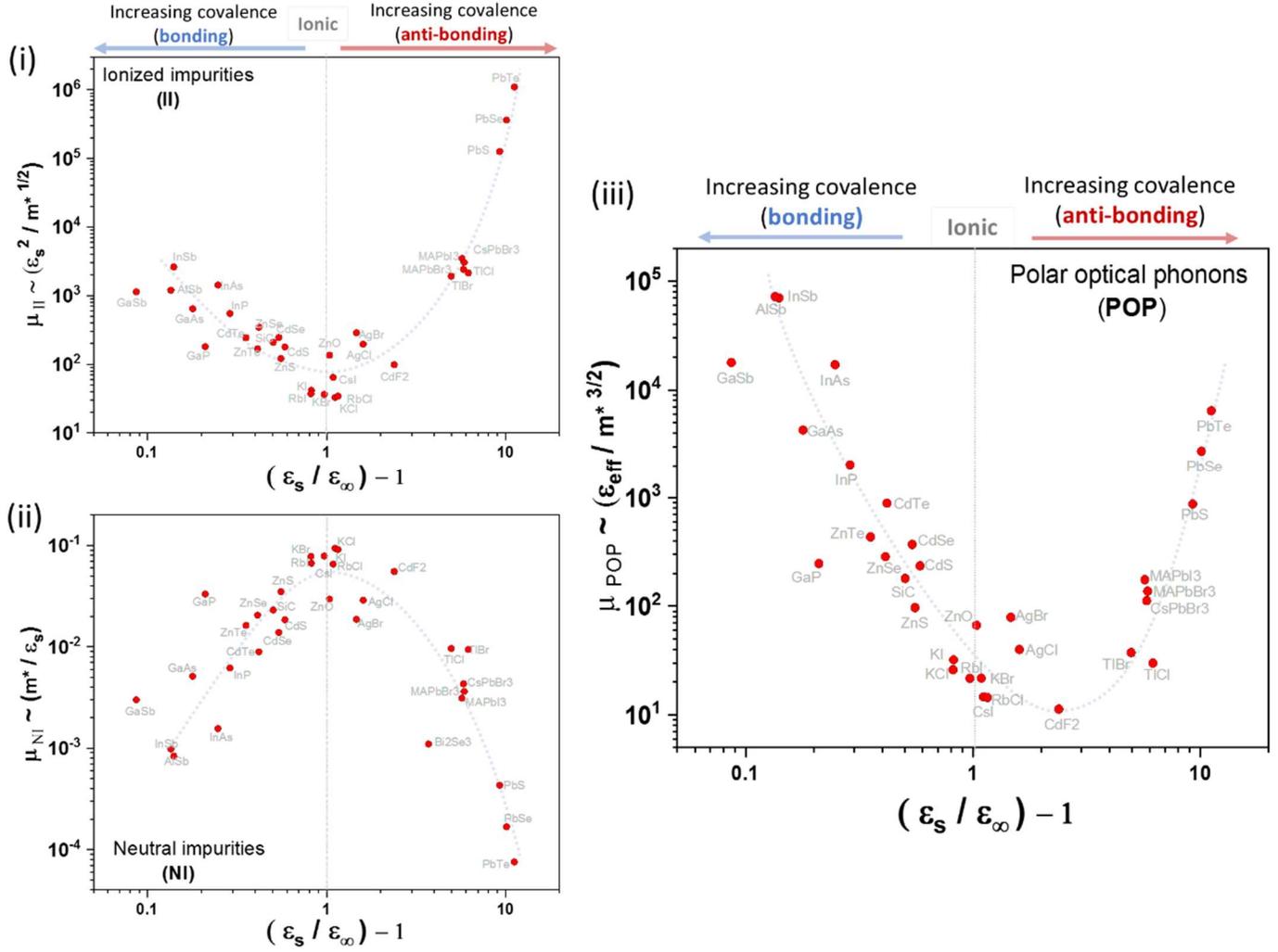

Figure 7: Log-log. plots of functions containing $m^*$, $\varepsilon_s$ and $\varepsilon_\infty$ that appear in red in Eq. 21 - Eq. 23, vs. RSP $\left(\frac{\varepsilon_s}{\varepsilon_\infty} - 1\right)$, reflecting the degree of covalence. The relevant scattering mechanism are (i) ionized and (ii) neutral impurities and (iii) polar optical phonons. Numerical values are calculated from the parameters in Table_S 1, Table_S 2 and Table_S 4. To understand the effect of the Debye temperature, we use the full proportionality in Eq. 23 with literature-derived longitudinal optical phonon frequencies ($\omega_{LO}$) (Table_S 5). **Fig. S5** shows that at room temperature (300K), the result is very similar to that presented in (iii) with only the $(\varepsilon_{eff}/(m^*)^{\frac{3}{2}})$ parameter, while for that at low temperature (10K), adding the expression $\frac{T^{\frac{1}{2}}}{\theta_D} \cdot \left(\exp\left(\frac{\theta_D}{T}\right) - 1\right)$ influences the correlation significantly.

It is important to note that these values are not the actual free charge mobilities, and prediction of the actual values requires the full analytical expressions of Eq. 21 - Eq. 23. Nevertheless, the predictive power of these plots is of some use, as, for example, when comparing temperature-dependent sets of data for GaAs[46] and PbSe[47]. According to their positions in Figure 7 (i) and (ii), at low temperatures (where defect-related scattering usually dominates) one should (and actually does) find that *charged* impurity scattering for GaAs (with $\mu \propto T^{+\frac{3}{2}}$) is dominant, while for PbSe scattering from *neutral* impurities (with $\mu \propto T^0$) dominates.

When low-defect density applications are required, such as in photovoltaics, or when temperatures are sufficiently high where phonons are highly active, for heteropolar compounds, scattering by polar optical phonons (*POP*) is usually the dominant scattering mechanism. Therefore, $\mu_{POP}$ defines, in principle, the highest possible mobility of heteropolar materials. With a very sharp dependence on RSP (Figure 7(iii)), $\mu_{POP}$ grows 3-4 orders of magnitude within one order of magnitude of RSP. As such, Figure 7(iii) reflects the fundamental mobility dependence with respect to the bond nature. Since the contribution of the mobility on the Debye temperature $\theta_D$ becomes dominant mostly at low temperatures (where $T < \theta_D$, which is usually < 300K), Figure 7(iii) provides a good estimate for the highest mobility a heteropolar material may reach around room temperature (see also **Fig. S5**).

### D. Non-radiative recombination:

For optoelectronic applications of semiconducting materials, the charge carrier lifetime is an important parameter that often controls optoelectronic functionality in combination with other parameters such as the absorption coefficient and the mobility. *Non-radiative recombination* (unlike radiative recombination) is always a parasitic process that has to be suppressed as much as possible to achieve high luminescence efficiency in LEDs and high open-circuit voltages in solar cells.[48,49] The theory of non-radiative recombination predicts that recombination depends both on the properties of the semiconductor and on that of the specific defect facilitating recombination. Thus, generic statements on recombination are difficult to make. However, because non-radiative recombination via defects is related to the dissipation of energy via emission of phonons, electron-phonon coupling will be important to understand non-radiative recombination, in a similar way to understanding transport.

Configuration coordinate diagrams illustrating the process of non-radiative transitions between two states (i.e. an electron state in the conduction band and a defect state) are presented in

**Fig. S8** and further explained in ESI - Section B. Based on this representation, for non-radiative recombination the Huang-Rhys factor, $S_{HR}$, is a crucial parameter that directly links to the non-radiative recombination rate. Ridley derives the Huang-Rhys factor of a generic defect[8] for the case of *deformation coupling* as:

Eq. 24) $\qquad S_{HR} \propto \left( \dfrac{D_{(p,o)}^2}{\omega^3 \cdot M_r} \right) I$

and for *polar coupling* as

Eq. 25) $\qquad S_{HR} \propto \left( \dfrac{1}{a_0 \cdot \omega \cdot \varepsilon_{eff}} \right) I$ .

The factor $I$ is a correction term[8] that includes the energy and charge state of traps and, thereby, takes into account that deep traps are the more localized ones, which increases the Huang-Rhys factor. $M_r$ is the reduced mass of the atomic oscillator, $a_0$ is the lattice constant (depends on lattice type) and $\omega$ is the frequency of the dominant phonon mode. The two parameters that we emphasize are the *optical* deformation potential constant, $D_{(p,o)}$, and the effective dielectric constant, $\varepsilon_{eff}$, which is the same as defined in Eq. 23 for $\mu_{POP}$.

Since $D_{(p,o)}$ may be very different from $D_p$ (the latter is the *acoustic* deformation potential), analysis of Eq. 24 and correlating $D_{(p,o)}$ and RSP is left to future work.

Despite its importance, there is only a limited amount of literature on measured or calculated values of $S_{HR}$. Examples for first principle calculations of $S_{HR}$ are presented in[50] dealing with charge recombination via defects in $Cu_2ZnSnS_4$ - a semiconductor investigated for applications in photovoltaics. In principle, the smaller the phonon energy and $S_{HR}$ are, the slower non-radiative recombination will be.

Here we focus on polar coupling of defects (Eq. 25). Since $S_{HR} \sim 1/\varepsilon_{eff}$, by correlating $1/\varepsilon_{eff}$ with our measure for covalence (as expressed by RSP), we find (Figure 8) that the smaller the covalence, the weaker the polar coupling of defects is, leading to slower non-radiative recombination. We note that (similar to the estimation of charge mobility) to estimate the actual non-radiative recombination probability, one will have to explicitly calculate $S_{HR}$ (from Eq. 24 and Eq. 25) and obtain parameters such as $\omega$, $D_{(p,o)}$ and $I$, which are not trivial to derive. Subsequently, one would have to use the Huang-Rhys factor to calculate the non-radiative recombination rates using models for multiphonon recombination such as the ones discussed in [51,52].

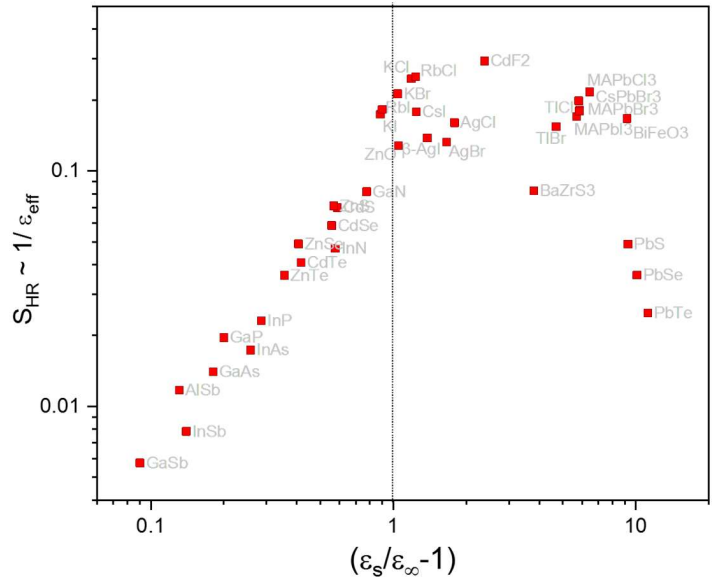

Figure 8: Log-log plot of $1/\varepsilon_{eff}$ (defined in Eq. 23) as function of RSP $\left(\frac{\varepsilon_s}{\varepsilon_\infty} - 1\right)$. The y-axis ($1/\varepsilon_{eff}$) is proportional to $S_{HR}$, thus represents the trend in the polar coupling regime. The higher the covalent nature of the material, the smaller $S_{HR}$ for polar coupling.

## IV.   SUMMARY

We studied novel approaches for identifying the *type* (i.e., attractive, 'bonding', or repulsive, 'anti-bonding') and *degree* (of orbital mixing, or the complementary properties: charge separation, or 'ionicity') of a covalent bond and its implications on fundamental electronic properties in heteropolar compounds. Using rationalization that was presented in the Model section, we use the relative structural polarizability (RSP) – an empirical parameter, defined as: $\left(\frac{\varepsilon_{ion}}{\varepsilon_{elect}}\right) \approx \left(\frac{\varepsilon_s}{\varepsilon_\infty} - 1\right)$ – as a reliable metric for the nature of the covalent bonding. To justify our relation of RSP to the nature of the bond, we found a good correlation between RSP and the 'classical' (Pauling) definitions for 'ionicity' (which is 1-'covalence'), as presented in Figure 3. This correlation can be used as an empirical calibration curve between the 'classical' and our definition for covalence. Unlike the classical definition, using RSP we can identify not only the degree, but also the type of covalence – a feature that was previously left for theoreticians, but that now can be approached experimentally. Furthermore, we identified a correlation between the type of the covalent nature and the bandgap-pressure coefficient, $\frac{\Delta E_g}{\Delta p}$ (see 'Model' section and further rationalization in ESI - Section A). Consequently, $\frac{\Delta E_g}{\Delta p}$ is suggested to be another empirical quantity for the type of covalence (at least qualitatively), as shown by its correlation with RSP (see Figure 4). Table I allows an experimentalist to estimate the nature of the bond.

The type of the covalent bond (attractive or repulsive) shows some chemical and structural trends. *Attractive* covalence is mostly found in highly covalent compounds that tend to form tetrahedrally coordinated (CN=4) structures. *Repulsive* covalence is mostly found in compounds with CN>4 (see **Fig. S1**(ii)) with relatively low 'covalence' ($f'_c < \sim 0.3$), in which the attractive electrostatic cohesive energy is sufficiently high to counterbalance the repulsive nature of the covalent bonding. Compounds with repulsive covalent nature may result from anti-bonding orbital mixing, but also due to 'geometric frustration'. The latter refers to situations where higher symmetry (leading to increasing entropy) is energetically more favorable than lowering enthalpy by symmetry-breaking. Examples for the latter effects are 2nd order Jahn-Teller effect or (as observed in compounds containing Sn, Pb, Bi, or Tl) ns²-lone pairs in the valence electron shell.

We used RSP to estimate implications of the bond nature on other fundamental properties that are important in semiconductors: electronic effective mass (Figure 5), the potential to screen electric fields (Figure 6), the mobility of free charge (Figure 7) and the excited charge recombination probability (Figure 8). We find:

- Materials with *attractive* covalent nature are more likely to respond to, or screen electric fields within a >THz frequency range, due to enhanced polarization of the electronic hard sphere.
- Materials with *repulsive* covalent nature are more likely to respond to, or screen electric fields within a <THz frequency range, due to additional structural polarizability.
- Electronic effective mass decreases with increase of the covalent nature *regardless* of the type of the covalent nature.

- Around room temperature and at the limit of low defect density (a limit that is specific for each material and depends on temperature), the highest (and most fundamental) mobility that a heteropolar compound can reach (i.e., $\mu_{POP}$) will increase with the degree of the covalent nature – regardless of the type.

- Scattering from ionized impurities (*II*) decreases with the increase of the degree of covalence, where highly ionic compounds are expected to have the lowest mobility. At the same time neutral impurities (*NI*) tend to limit the mobility of highly ionic materials the least.

- Polar coupling of defects reduces with the degree of covalent bonding (regardless of the type), leading to slower non-radiative recombination via polar coupling.

The new observables (RSP and $\frac{\Delta E_g}{\Delta p}$) for defining the nature of the covalent bond should allow estimating the potential of new (and old) compounds as (opto)electronic materials, as well as help find new correlations with other applicative properties (e.g., electrostriction – see comment *xi*). It is clear that the proposed model has limitations and further work may allow its application to a broader set of materials and, consequently, help improve the physical and chemical intuitive understanding of materials with respect to their bond nature.

## V.     ACKNOWLEDGMENTS

We thank Omer Yaffe and Leeor Kronik (Weizmann Institute of Science), Douglas Fabini (MPI FKF), Aaron Walsh (Imperial College London) and Umesh Waghmare and Shashank Chaturvedi (JNCASR) for discussions that were essential for the completion of this manuscript.

Electronic Supplementary Information (ESI)

# *Type* and *Degree* of Covalence:
# Empirical Derivation and Implications

*Yevgeny Rakita[1\*], Thomas Kirchartz[2,3], Gary Hodes[1] , David Cahen[1]*

\* *yevgev@gmail.com*

*Present Address:  Applied Physics and Applied Mathematics, Columbia University, 500 w 120st Mudd #1105,*
*10025, New York, NY, USA  ;  yr2369@columbia.edu*

1. Department of Materials and Interfaces, Weizmann Institute of Science, Rehovot, 76100, Israel
2. IEK5-Photovoltaics, Forschungszentrum Jülich, 52425 Jülich, Germany
3. Faculty of Engineering and CENIDE, University of Duisburg–Essen, Carl-Benz-Strasse 199, 47057 Duisburg, Germany

Table_S 1: Experimentally-derived: bulk moduli (*B*); dEg/dP at lowest pressure values available of $\frac{\Delta E_g}{\Delta p}$ ; calculated deformation potential (*D_P*) using Eq. 13. Values refer to measurements at 300 K. The relevant references are mention to the right of each value. The way we have been extracting $\frac{\Delta E_g}{\Delta p}$ is described in **Fig. S7** (ESI – Section A).

| | Structure (@ RT ) | Coordination number (CN) | B | Ref. | dEg/dP | Ref. | D_P (Eq. 13) |
|---|---|---|---|---|---|---|---|
| MAPbI₃ | Perovskite | 6,12 | 13.9 | [53] | -0.088 | [54] | 1.22 |
| CsPbBr₃ | Perovskite | 6,12 | 15.5 | [53] | -0.049 | [55] | 0.76 |
| MAPbBr₃ | Perovskite | 6,12 | 15.6 | [53] | -0.042 | [56] | 0.66 |
| MAPbCl₃ | Perovskite | 6,12 | 25 | [57] | -0.078 | [58] | 1.95 |
| BaZrS₃ | Perovskite | 6,12 | 75 | [59] | -0.016 | [59] | 1.20 |
| BiFeO₃ | Perovskite | 6,12 | 122 | [60] | -0.058 | [60] | 7.08 |
| TlCl | BCC (CsCl-like) | 8 | 16 | [61] | -0.134 | [62] | 2.14 |
| TlBr | BCC (CsCl-like) | 8 | 12 | [61] | -0.139 | [62] | 1.67 |
| AgCl | Rocksalt | 6 | 43.3 | [63] | -0.019 | [62] | 0.74 |
| AgBr | Rocksalt | 6 | 39.9 | [63] | -0.015 | [63] | 0.60 |
| **β** −AgI | Wurtzite | 4 | 33.8 | [63] | -0.061 | [63] | 2.94 |
| PbTe | Rocksalt | 6 | 46 | [64] | -0.074 | [65] | 3.41 |
| PbSe | Rocksalt | 6 | 54 | [64] | -0.086 | [65] | 4.64 |
| PbS | Rocksalt | 6 | 62 | [64] | -0.060 | [65] | 3.72 |
| CdTe | Zincblende | 4 | 45 | [17] | 0.0760 | [17] | -3.42 |
| CdSe | Wurtzite | 4 | 53 | [17] | 0.058 | [17] | -3.07 |
| CdS | Wurtzite | 4 | 62 | [17] | 0.044 | [17] | -2.73 |
| GaSb | Zincblende | 4 | 56 | [17] | 0.140 | [17] | -7.84 |
| GaAs | Zincblende | 4 | 75 | [17] | 0.085 | [17] | -6.38 |
| GaP | Zincblende | 4 | 88 | [17] | 0.097 | [17] | -8.54 |
| GaN | Wurtzite | 4 | 205 | [17] | 0.04 | [17] | -8.20 |
| InSb | Zincblende | 4 | 48 | [17] | 0.130 | [17] | -6.24 |
| InAs | Zincblende | 4 | 58 | [17] | 0.114 | [17] | -6.61 |
| InP | Zincblende | 4 | 71 | [17] | 0.080 | [17] | -5.68 |
| InN | Wurtzite | 4 | 148 | [17] | 0.020 | [17] | -2.96 |
| Si | Diamond | 4 | 97.9 | [17] | 0.030 | [66] | -2.94 |
| Ge | Diamond | 4 | 68.9 | [17] | 0.055 | [66] | -3.79 |

Table_S 2: $\varepsilon_\infty$ and $\varepsilon_s$ values that are based on experimental data. The references from which the data are taken are given in the right column. Calculated $\frac{\varepsilon_{ionic}}{\varepsilon_{elect}}$ and $\varepsilon^*$ values are presented. Values refer to 300 K, unless stated otherwise.

| | $\varepsilon_\infty$ | $\varepsilon_s$ | Ref. | $RSP = \frac{\varepsilon_{ionic}}{\varepsilon_{elect}} = \left(\frac{\varepsilon_s}{\varepsilon_\infty} - 1\right)$ |
|---|---|---|---|---|
| MAPbI$_3$ | 5.0 | 33.5 | [67] | 5.70 |
| CsPbBr$_3$ | 4.3 | 29.3 | [68] | 5.81 |
| MAPbBr$_3$ | 4.7 | 32.3 | [67] | 5.87 |
| MAPbCl$_3$ | 4.0 | 29.8 | [67] | 6.45 |
| BaZrS$_3$ | 9.6 | 46.0 | [69] | 3.79 |
| BiFeO$_3$ | 5.4 | 55.0 | [70] | 9.19 |
| Bi$_2$Se$_3$ | 29.0 | 113 | [71] | 2.90 |
| TlCl | 4.76 | 32.60 | [72] | 5.85 |
| TlBr | 5.34 | 30.40 | [72] | 4.69 |
| AgCl | 4.0 | 11.15 | [72] | 1.79 |
| AgBr | 4.7 | 12.50 | [72] | 1.66 |
| $\beta$-AgI | 4.2 | 10 | [73,74] | 1.38 |
| PbTe | 36.9 | 450.0 | [72] | 11.20 |
| PbSe | 25.2 | 280.0 | [72] | 10.11 |
| PbS | 18.5 | 190.0 | [72] | 9.27 |
| CdTe | 7.2 | 10.2 | [72] | 0.42 |
| CdSe | 6.1 | 9.50 | [72] | 0.56 |
| CdS | 5.3 | 8.42 | [72] | 0.59 |
| GaSb | 14.4 | 15.7 | [72] | 0.09 |
| GaAs | 10.9 | 12.87 | [72] | 0.18 |
| GaP | 8.5 | 10.20 | [72] | 0.20 |
| GaN | 5.35 | 9.50 | [39] | 0.78 |
| InSb | 15.7 | 17.9 | [72] | 0.14 |
| InAs | 11.8 | 14.84 | [72] | 0.26 |
| InP | 9.6 | 12.34 | [72] | 0.29 |
| InN | 7.8 | 12.30 | [39] | 0.58 |
| **KBr** | 2.4 | 4.90 | [72] | 1.04 |
| **KCl** | 2.2 | 4.81 | [72] | 1.19 |
| **KI** | 2.7 | 5.09 | [72] | 0.89 |
| **RbCl** | 2.2 | 4.92 | [72] | 1.24 |
| **RbI** | 2.6 | 4.94 | [72] | 0.90 |
| **CsI** | 3.1 | 6.95 | [72] | 1.24 |
| **ZnS** | 5.1 | 8.0 | [72] | 0.57 |
| **ZnSe** | 5.9 | 8.3 | [72] | 0.41 |
| **ZnTe** | 7.3 | 9.9 | [72] | 0.36 |
| **ZnO** | 4.0 | 8.2 | [72] | 1.05 |
| **AlSb** | 9.9 | 11.2 | [72] | 0.13 |
| **CdF$_2$** | 2.4 | 8.1 | [72] | 2.38 |
| **SiC** | 6.7 | 10.0 | [72] | 0.49 |
| **Si** | 11.90 | 11.90 | [75] | 0 |
| **Ge** | 16.20 | 16.20 | [75] | 0 |

Table_S 3: Parameters, used to extract the 'covalence ' using Pauling's 'resonating bond' model[2], following Eq. 2. $M$ and $N$ are for the room-temperature phases. $X_A$ and $X_B$ are taken from three different sources (see top row for references) and plotted against RSP $\left(\frac{\varepsilon_s}{\varepsilon_\infty} - 1\right)$ in Fig. S2. $X_A$ and $X_B$ for perovskite structures were related to the B cation and the X anion, respectively, since these are the atoms closest to each other in perovskite structures

| Material | Anion valence | Effective CN | Pauling[76] | | Phillips[2] | | Lang and Smith[29] | |
|---|---|---|---|---|---|---|---|---|
| | N | M | $X_A$ | $X_B$ | $X_A$ | $X_B$ | $X_A$ | $X_B$ |
| InN | 3 | 4 | 1.78 | 3.04 | 0.99 | 3 | 1.26 | 2.82 |
| InP | 3 | 4 | 1.78 | 2.19 | 0.99 | 1.64 | 1.26 | 2.05 |
| GaN | 3 | 4 | 1.81 | 3.04 | 1.13 | 3 | 1.31 | 2.82 |
| GaP | 3 | 4 | 1.81 | 2.19 | 1.13 | 1.64 | 1.31 | 2.05 |
| GaSb | 3 | 4 | 1.81 | 2.05 | 1.13 | 1.31 | 1.31 | 1.73 |
| InAs | 3 | 4 | 1.78 | 2.18 | 0.99 | 1.57 | 1.26 | 1.95 |
| GaAs | 3 | 4 | 1.81 | 2.18 | 1.13 | 1.57 | 1.31 | 1.95 |
| InSb | 3 | 4 | 1.78 | 2.05 | 0.99 | 1.31 | 1.26 | 1.73 |
| CdTe | 2 | 4 | 1.69 | 2.1 | 0.83 | 1.47 | 2.06 | 2.01 |
| CdSe | 2 | 4 | 1.69 | 2.55 | 0.83 | 1.79 | 2.06 | 2.3 |
| CdS | 2 | 4 | 1.69 | 2.58 | 0.83 | 1.87 | 2.06 | 2.49 |
| MAPbBr3 | 1 | 6 | 1.87 | 2.96 | 1.09 | 2.01 | 1.51 | 2.67 |
| AgI | 1 | 4 | 1.93 | 2.66 | 0.57 | 1.63 | 1.92 | 2.32 |
| AgBr | 1 | 6 | 1.93 | 2.96 | 0.57 | 2.01 | 1.92 | 2.67 |
| CsPbBr3 | 1 | 6 | 1.87 | 2.96 | 1.09 | 2.01 | 1.51 | 2.67 |
| AgCl | 1 | 6 | 1.93 | 3.16 | 0.57 | 2.1 | 1.92 | 2.95 |
| MAPbI3 | 1 | 6 | 1.87 | 2.66 | 1.09 | 1.63 | 1.51 | 2.32 |
| TlBr | 1 | 8 | 1.62 | 2.96 | 0.94 | 2.01 | 1.34 | 2.67 |
| MAPbCl3 | 1 | 6 | 1.87 | 3.16 | 1.09 | 2.1 | 1.51 | 2.95 |
| TlCl | 1 | 8 | 1.62 | 3.16 | 0.94 | 2.1 | 1.34 | 2.95 |
| PbTe | 2 | 6 | 1.87 | 2.1 | 1.09 | 1.47 | 1.51 | 2.01 |
| PbS | 2 | 6 | 1.87 | 2.58 | 1.09 | 1.87 | 1.51 | 2.49 |
| PbSe | 2 | 6 | 1.87 | 2.55 | 1.09 | 1.79 | 1.51 | 2.3 |
| BiFeO3 | 2 | 6 | 1.83 | 3.44 | 1.2 [a] | 3.5 | 1.77 | 3.39 |
| BaZrS3 | 2 | 6 | 1.33 | 2.58 | 0.6 [a] | 1.87 | 1.57 | 2.49 |
| KBr | 1 | 6 | 0.82 | 2.66 | 0.4 [a] | 2.01 | 1 | 2.67 |
| KCl | 1 | 6 | 0.82 | 3.16 | 0.4 [a] | 2.1 | 1 | 2.95 |
| KI | 1 | 6 | 0.82 | 2.66 | 0.4 [a] | 1.63 | 1 | 2.32 |
| RbCl | 1 | 6 | 0.82 | 3.16 | 0.4 [a] | 2.1 | 0.96 | 2.95 |
| RbI | 1 | 6 | 0.82 | 2.66 | 0.4 [a] | 1.63 | 0.96 | 2.32 |
| CsI | 1 | 8 | 0.79 | 2.66 | 0.4 [a] | 1.63 | 0.89 | 2.32 |
| ZnS | 2 | 4 | 1.65 | 2.58 | 0.91 | 1.87 | 2.16 | 2.49 |
| ZnSe | 2 | 4 | 1.65 | 2.55 | 0.91 | 1.79 | 2.16 | 2.3 |
| ZnTe | 2 | 4 | 1.65 | 2.1 | 0.91 | 1.47 | 2.16 | 2.01 |
| ZnO | 2 | 4 | 1.65 | 3.44 | 0.91 | 3.5 | 2.16 | 3.39 |
| AlSb | 3 | 4 | 1.61 | 2.05 | 1.18 | 1.31 | 1.31 | 1.73 |
| CdF2 | 2 | 6 | 1.69 | 3.98 | 0.83 | 4 | 2.06 | 4 |

[a] These values are not available explicitly. Therefore, they are assumptions based on rational extrapolations.

Table_S 4: $m^*$ values that are based on experimental data. Most values are derived from cyclotron resonance or Faraday rotation experiments except those for HaPs, where magneto-resistance experiments are used.

| | Effective mass, $m^*$ | Ref. |
|---|---|---|
| **MAPbI₃** | 0.104 | [77] |
| **MAPbBr₃** | 0.117 | [77] |
| **CsPbBr₃** | 0.126 | [78] |
| **KBr** | 0.369 | [72] |
| **KCl** | 0.434 | [72] |
| **KI** | 0.325 | [72] |
| **RbCl** | 0.432 | [72] |
| **RbI** | 0.368 | [72] |
| **CsI** | 0.42 | [72] |
| **TlBr** | 0.315 | [72] |
| **TlCl** | 0.33 | [72] |
| **AgBr** | 0.215 | [72] |
| **AgCl** | 0.297 | [72] |
| **GaAs** | 0.0657 | [72] |
| **GaSb** | 0.047 | [72] |
| **GaP** | 0.338 | [72] |
| **InAs** | 0.023 | [72] |
| **InSb** | 0.015 | [72] |
| **CdS** | 0.155 | [72] |
| **CdSe** | 0.13 | [72] |
| **CdTe** | 0.091 | [72] |
| **ZnS** | 0.28 | [72] |
| **ZnSe** | 0.171 | [72] |
| **ZnTe** | 0.16 | [72] |
| **ZnO** | 0.24 | [72] |
| **PbS** | 0.082 | [72] |
| **PbSe** | 0.047 | [72] |
| **PbTe** | 0.034 | [72] |
| **AlSb** | 0.011 | [72] |
| **InP** | 0.076 | [72] |
| **CdF₂** | 0.45 | [72] |
| **SiC** | 0.23 | [72] |
| **Si** | 0.19 | [72] |
| **Ge** | 0.08 | [72] |

Table_S 5: $\omega_{LO}$ values that are based on experimental data and related $\theta_D = \omega_{LO}[Hz] \cdot \frac{h}{k_B}$, where $h$ and $k_B$ are Planck's and Boltzmann's coefficients.

| | $\omega_{LO}$ (@ ~300) [cm$^{-1}$] | Ref. | $\theta_D$ [K] |
|---|---|---|---|
| **MAPbI3** | 133 | [67] | 191 |
| **CsPbBr3** | 136 | [68] | 196 |
| **MAPbBr3** | 166 | [67] | 239 |
| **MAPbCl3** | 225 | [67] | 324 |
| **TlCl** | 164.9 | [72] | 237 |
| **TlBr** | 114.3 | [72] | 164 |
| **AgCl** | 176.8 | [72] | 254 |
| **AgBr** | 129.9 | [72] | 187 |
| **PbTe** | 110.0 | [72] | 158 |
| **PbSe** | 146.7 | [72] | 211 |
| **PbS** | 213.7 | [72] | 307 |
| **CdTe** | 168.0 | [72] | 242 |
| **CdSe** | 213.4 | [72] | 307 |
| **CdS** | 303.5 | [72] | 437 |
| **GaSb** | 240.3 | [72] | 346 |
| **GaAs** | 291.9 | [72] | 420 |
| **GaP** | 403.0 | [72] | 580 |
| **InSb** | 197.2 | [72] | 284 |
| **InAs** | 241.8 | [72] | 348 |
| **InP** | 345.0 | [72] | 496 |
| **Si** | 449.0 | [72] | 646 |
| **Ge** | 263.0 | [72] | 378 |
| **KBr** | 163.2 | [72] | 235 |
| **KCl** | 210.0 | [72] | 302 |
| **KI** | 140.6 | [72] | 202 |
| **RbCl** | 173.0 | [72] | 249 |
| **RbI** | 103.9 | [72] | 149 |
| **CsI** | 90.5 | [72] | 130 |
| **ZnS** | 352.0 | [72] | 506 |
| **ZnSe** | 246.0 | [72] | 354 |
| **ZnTe** | 206.0 | [72] | 296 |
| **ZnO** | 519.0 | [72] | 747 |
| **CdF2** | 380.0 | [72] | 547 |

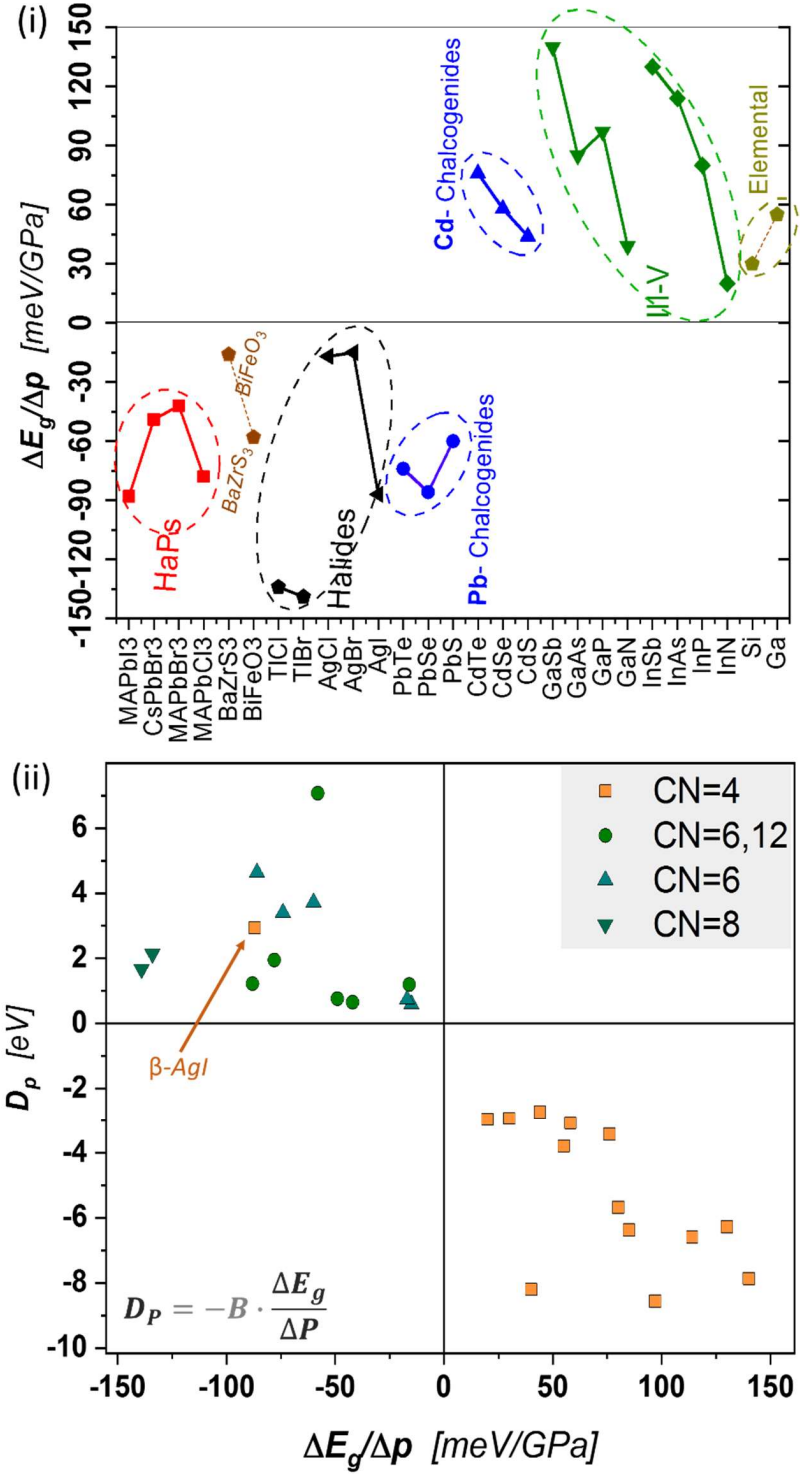

**Fig. S1**: (i) $\frac{\Delta E_g}{\Delta p}$ for different materials based on the values that are presented in Table_S 1. (ii) The Correlation between the coordination number (CN), $\frac{\Delta E_g}{\Delta p}$ and $D_p$ showing that among the 25 heteropolar and 2 homopolar chosen materials, all of the materials with CN=4 are with $D_p < 0$ (and $\frac{\Delta E_g}{\Delta p} > 0$), while all of those with CN>4 are with $D_p > 0$ (and $\frac{\Delta E_g}{\Delta p} < 0$), with only one exception – $\beta$-AgI, that turns to $\alpha$-AgI (rocksalt structure (CN=6)) at ~145 °C, like the other Ag-halides.

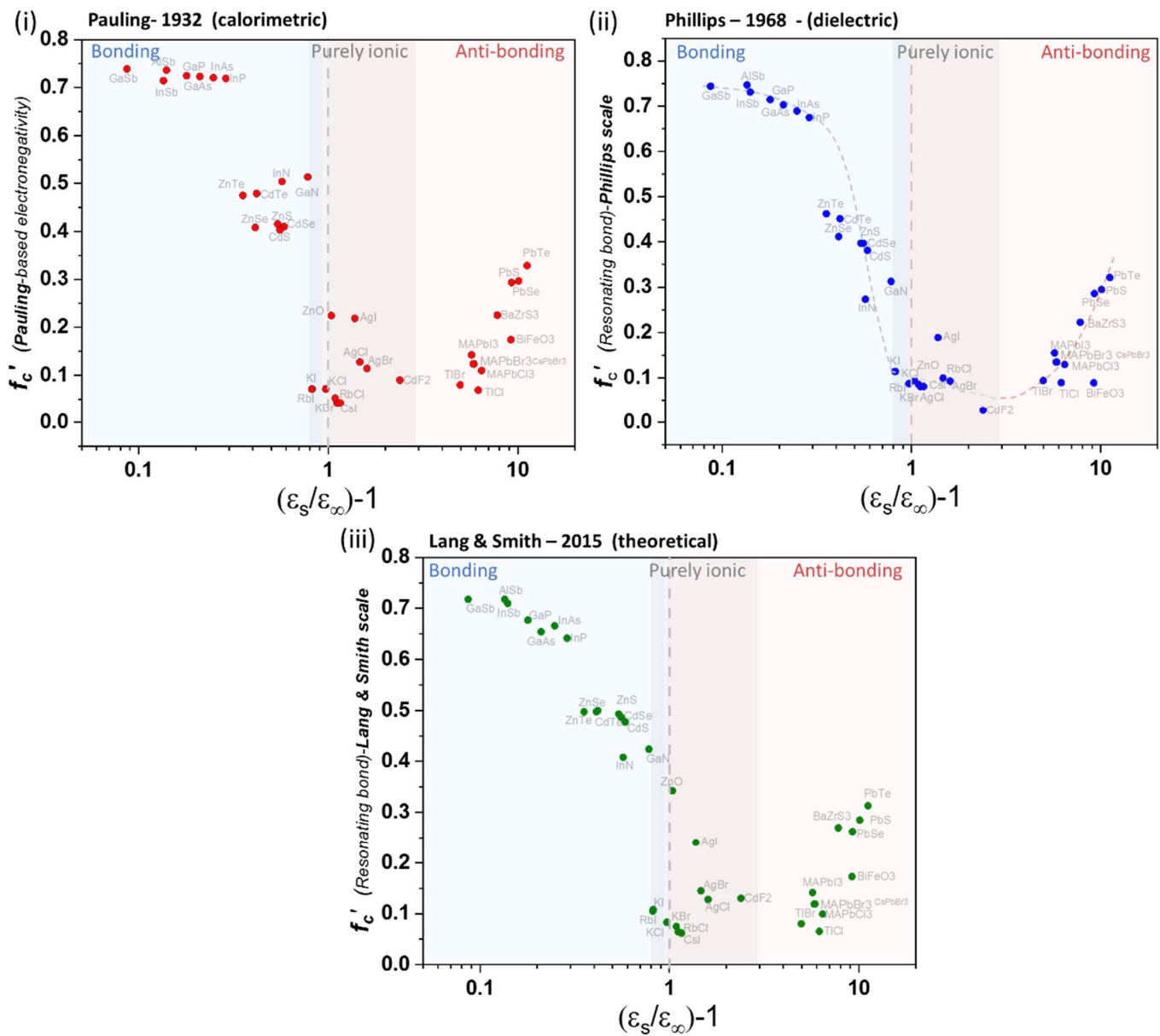

**Fig. S2:** Semi-log correlation between the Pauling definition for 'covalence' (Eq. 2) with respect to our proposed RSP (Eq. 19). Dielectric and electronegativity values are summarized in Table_S 2 and Table_S 3. Different values of electronegativity are based on (i) the original values of Pauling (1932)[1] that are based on calorimetry (bond energy), (ii) Phillips' (1968)[2,30] definition that was corrected by adding the Thomas-Fermi screening factor and based on dielectric properties[30], and (iii) the latest corrected values by Lung & Smith (2015)[29]. For Perovskites structures we used the electronegativity values for the B cation and X anion (in an ABX$_3$ composition), as the shortest, and thus most dominating backbone of the structure.

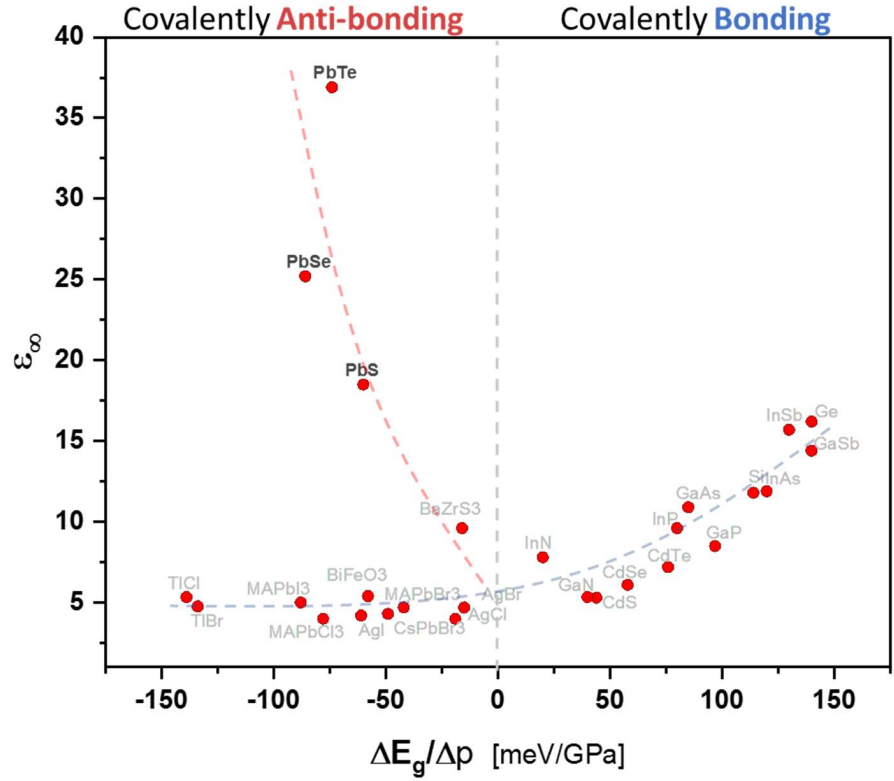

**Fig. S3**: A similar figure to Figure 6, but to a further scale at the y-axis (a lin.-lin. correlation between the bandgap-pressure coefficient $\frac{\Delta(E_g)}{\Delta p}$ the *electronic* polarizability (EP - $\varepsilon_\infty$)) that includes the exception of Pb-chalcogenides, that have much higher $\varepsilon_\infty$ than the rest of the materials. The origin for the exception is unclear to us. See Table_S 1 and Table_S 2 for the numerical values.

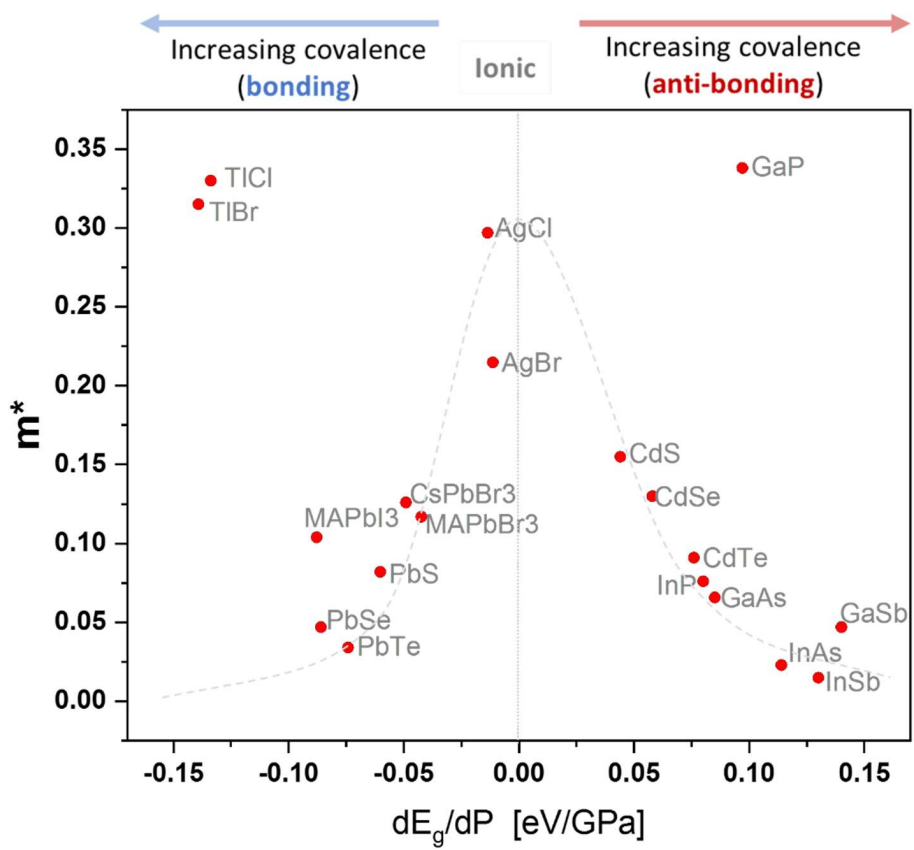

**Fig. S4:** Log-lin. plot of the electronic effective mass, $m^*$, with respect to $\frac{\Delta E_g}{\Delta p}$. The numerical values are experimentally-derived data – see Table_S 1 and Table_S 4.

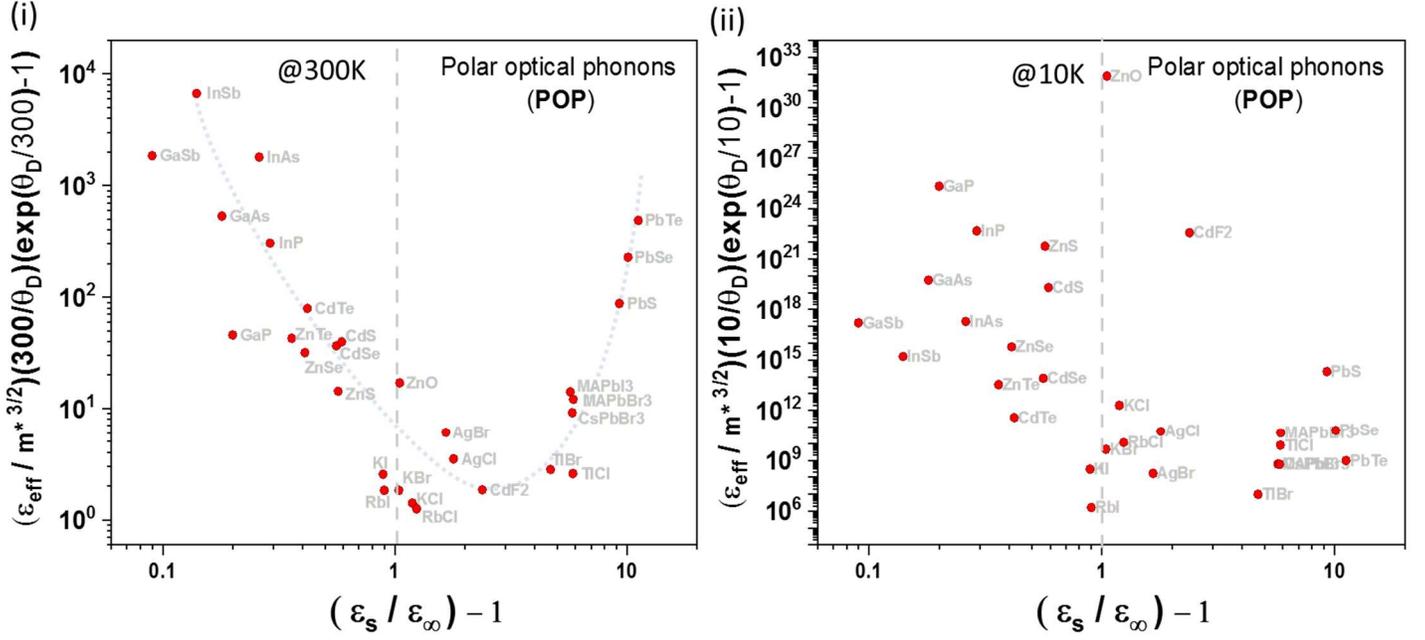

**Fig. S5:** Log-log. correlation between the proportion relation for $\mu_{POP}$ (following Eq. 23) against RSP using literature derived $m^*$ (Table_S 4) , $\varepsilon_s$, $\varepsilon_\infty$ (Table_S 2) and longitudinal optical phonon frequencies ($\omega_{LO}$) from which we derived $\theta_D$ (see values and derivation in Table_S 5). At room temperature (300K), the result is very similar to that presented in Figure 7(iii) with only the $(\varepsilon_{eff}/(m^*)^{\frac{3}{2}})$ parameter, while for that at low temperature (10K), the addition of the expression $\frac{T^{\frac{1}{2}}}{\theta_D} \cdot \left(\exp\left(\frac{\theta_D}{T}\right) - 1\right)$, influences the correlation significantly.

**Section A:**

**_About deformation potential and its relation to nature of the covalent bond_**

With compression (or dilation) the potential fields change, leading to differences in both VBM and CBM and, similarly to $D_p$, the band energy change with atomic spacing can be defined as $D_{(VBM)}$ or $D_{(CBM)}$, respectively. $D_p$ can then be defined as: $\boldsymbol{D_p} = D_{(VBM)} - D_{(CBM)}$ so when $D_{(CBM)} > D_{(VBM)}$, then $\boldsymbol{D_p} < 0$, and vise versa. The case of $D_{(CBM)} > D_{(VBM)}$ is commonly studied in textbooks, since textbooks usually refer to the classical cases of of zincblende and wurtzite structured semiconductors (CN=4), such as Si and GaAs, as can also be found in a study of Wei and Zunger (1999).[17] A common way to rationalize $D_{(CBM)} > D_{(VBM)}$ is from electronic band energies in real space, i.e., vs. interatomic spacing (outcome of tight binding modeling - see for example chapter 4 in ref. [28]). Similar schematic representation is drawn in Figure 1(ii). For CN=4 systems, where VBM is constructed out of 'bonding' ($\sim\sigma[sp^3\text{-}sp^3]$) orbitals, its energy will decrease (or barely change), leading to a negative or very small $D_{(VBM)}$. In contrary, CBM, which is antibonding ($\sim\sigma^*[sp^3\text{-}sp^3]$) and represents strong repulsion between the two $\sigma^*$ orbitals, will increase in with compression, resulting in a positive $D_{(CBM)}$. Following Figure 1(ii), the overall result should leads to $\boldsymbol{D_p} < 0$.

In rocksalt Pb-chalcogenides, unlike zincblende Cd-chalcogenides, although they are both II-VI compounds with very similar bulk moduli and _absolute_ response to pressure, i.e. $\left|\frac{\Delta E_g}{\Delta p}\right|$, (Fig. S1(i)), the algebraic sign of $\boldsymbol{D_p}$ (due to $\frac{dE_g}{dP}$) is _opposite_. Following Wei and Zunger (1997),[19] unlike in II-VI and III-V CN=4 SCs, in Pb-chalcogenide the VBM (which occurs at the L point and not at the $\Gamma$ as for CN=4 SCs) is constructed of $6s_{(Pb)}$-$np_{(S,Se,Te)}$ repulsive (i.e., antibonding) orbitals that increase the VBM energy upon compression (i.e. $D_{(VBM)} > 0$). The CBM, is constructed mostly of $6p_{(Pb)}$ ($\sim$non-bonding) can construct also repulsive $6p_{(Pb)}$-$ns_{(S,Se,Te)}$ coupling; however, spin-orbit coupling strongly reduces the CBM energy (as also shown for HaPs)[79]. Although the valence $p$ orbitals of the chalcogenides decrease from S to Te, and the conduction $s$ orbitals increase from S to Te, the strong repulsive interactions with Pb at the VBM and spin-orbit coupling at the CBM, leads to strong suppression of the bandgap width. The different covalent coupling strengths lead to anomalous bandgap ordering, i.e., $E_g(PbS) > E_g(PbTe) > E_g(PbSe)$, unlike the relevant orbital energies of the.[19] Overall, repulsive covalent bonding should lead to $\boldsymbol{D_p} < 0$.

HaPs are also known to have 'anti-bonding' VBM,[21] but unlike Pb-chalcogenides we know that the actual bandgap is systematically decreasing when moving from Cl to Br to I, which suggests that repulsive nature of $m(s_{(Pb,Sn)})$-$n(p_{(Cl,Br,I)})$ is less dominant than that of Pb-chalcogenides. Following calculations preformed in Fabini et al.,[80] it was shown that with increasing pressure, the width of the VBM changes severely while the CBM is remained almost unchanged (see also **Fig. S6**), leading to an overall similar picture to what our illustration in Figure 1(ii).

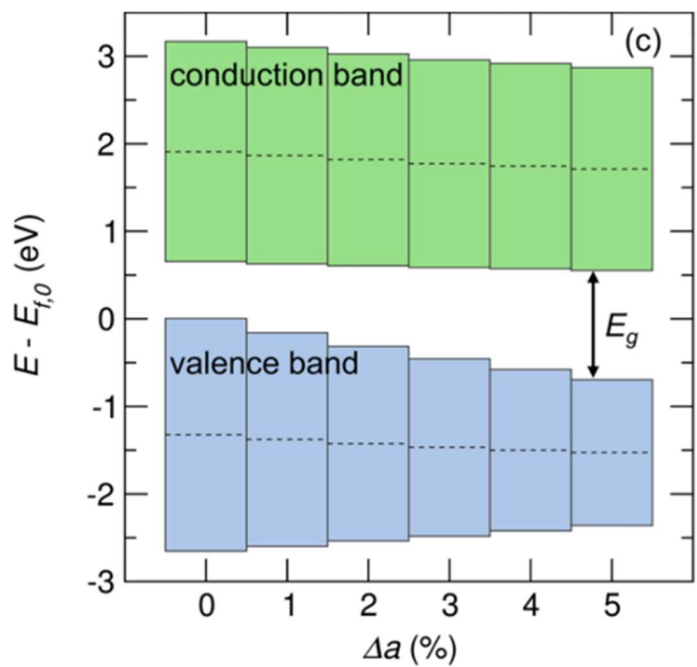

**Fig. S6**: A figure and its caption that are adopted (used with permission-<span style="color:blue">REQUEST SENT</span>) from Fabini et al.[80] that supports the rationalization behind Figure 2 (ii) for 'anti-bonding' hybridization and development with interatomic distance: 'Evolution of the CsSnBr₃ valence band and conduction band edges with lattice expansion, showing the simultaneous narrowing and stabilization of both bands due to weakened antibonding interactions. Band centers are indicated by dashed lines, and all are plotted on a common energy scale by aligning the energy of dispersionless Cs states. This is contrary to the bandgap evolution of tetrahedral semiconductors and of isostructural CsCaBr₃, which lacks an s² lone pair.'

With regard to 'degree of covalence', for example in perovskites the degree of orbital overlap is of critical importance,[81] where even the angles formed between two corner-sharing octahedra immediately affects its bandgap. Following **Fig. S1**(i), Ga, who is energetically closer than In to nitrogen, possess a larger $\left|\frac{\Delta E_g}{\Delta p}\right|$. From another perspective, BiFeO₃ is an example for covalently-matched material leading to a relatively large $\frac{\Delta E_g}{\Delta p}$. Different Fe-O and Ni-O perovskites are known to have quite an extensive covalent match that leads to unique effects (such as ligand-holes and metal-to-insulator phase transitions).[43,44] Generally, larger atoms tend to have bulkier electronic cloud, which tend to add covalent contribution to the bond. One can try to be convinced by comparing Tl-halides with respect to Ag-halides, or when comparting Ga- or In-pnictides with increasing weight of the pnictide (N → Sb). Following these examples, we find $\left|\frac{\Delta E_g}{\Delta p}\right|$ as a parameter that can serve as a rough estimation to the degree of covalence of materials, meaning that higher $\left|\frac{\Delta E_g}{\Delta p}\right|$ indicates a greater covalent contribution to the bond. To some extent, this is the case for 'bonding' covalent materials, as shown in Figure 4 but much less for 'anti-bonding' ones. Further discussion is found in the main text –see 'Model' section.

Nevertheless, an exact degree of covalence will require additional information on, for example, the type and energy position of atomic/molecular valence orbitals, the interatomic distance and structural symmetry. Solid-state NMR is suggested as another experimental tool to get more information on the electronic density and covalence of bonds; however, theoretical support is definitely necessary. The algebraic sign will indicate the *type* of the covalent interaction: *positive* is for attracting interactions between orbitals (i.e., 'bonding'); *negative* is for repulsive interactions (i.e., 'anti-bonding'). Following the suggestion for complimentary solid-state NMR experiments, we find that the trend in the chemical shift of $Cd^{113}$ in Cd-chalcogenides[82] is opposite to the trend found for $Pb^{207}$ in $APbX_3$ [83] when probing heavier anions. In the $APbX_3$ HaPs, where the VBM is assumed to be 'anti-bonding', the electron density around $Pb^{2+}$ is *decreased* with the assumed degree of covalence ($Cl \rightarrow I$), while for $Cd^{2+}$ is *increased* with the assumed degree of covalence between the hybridized orbitals ($S \rightarrow Te$), as expected from a bonding orbital. The chemical shifts of $Pb^{207}$ in Pb-chalcogenides[84] are inconsistent (Pb electron density: PbTe>PbS>PbSe) and may be related to the contribution of free-charges (due to the very low bandgap - cf. Knight's shift)[85] or the degree of covalence between the atomic orbitals[19]. Besides $\frac{\Delta E_g}{\Delta p}$, which is presented here, other (theoretically-supported) empirical techniques, such as solid-state NMR, are highly encouraged. Following the complexity and reliability of the mentioned methods for assessment of the bond nature, in this paper we suggest to use RSP as an easier and more reliable measure for the type and degree of covalence in heteropolar compounds.

***Ways to measure $D_p$*** in a way that it will give us an estimate on the type of the covalent interactions at the VBM level is by measuring the slope for $\frac{\Delta E_g}{\Delta p}$ or $\frac{\Delta E_g}{\Delta T}$ multiplied by ($-B$) or ($1/\alpha_T$). To derive $\frac{\Delta E_g}{\Delta p}$ or $\frac{\Delta E_g}{\Delta T}$ for $D_p$ we extrapolated the presented experimental points (see examples for $MAPbI_3$ and CdTe in **Fig. S7**. At temperatures around 300K and pressure near atmospheric pressure. For the determination of $E_g$ we usually preferred 'absorption' data rather than from 'photoluminescence' (PL) data.

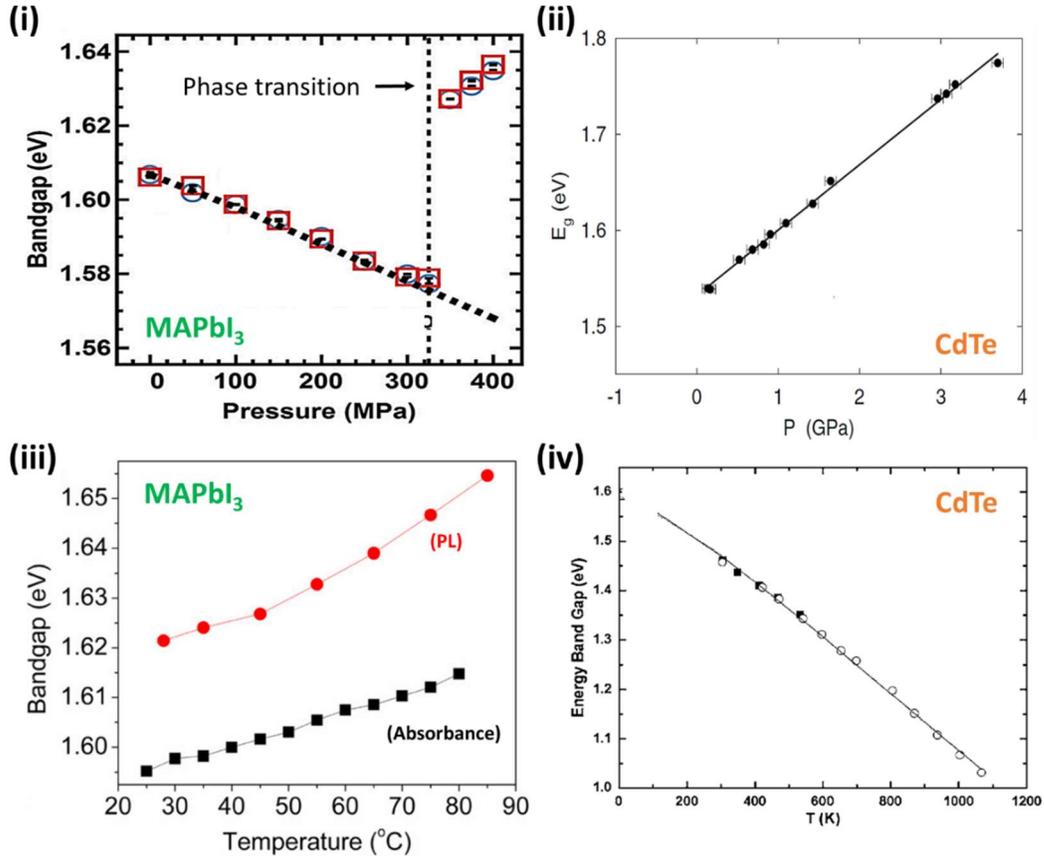

**Fig. S7**: Experimental data (used with permission-<span style="color:blue">REQUEST SENT</span>) for: $E_g$ (via optical absorbance) vs. pressure at 300K for (**i**) MAPbI$_3$[54] and (**ii**) CdTe[86]; $E_g$ (via optical absorbance (or PL peak for MAPbI$_3$)) vs. temperature at atmospheric pressure for (**iii**) MAPbI$_3$[87] and (**iv**) CdTe[88]. By using the material's parameters:[53] $B(MAPbI_3) = 13.9 \, [GPa]$, $B(CdTe) = 45 \, [GPa]$, $\alpha_T(MAPbI_3) = 1.3 \cdot 10^{-4} \left[\frac{1}{K}\right]$ and $\alpha_T(CdTe) = 1.8 \cdot 10^{-5} \left[\frac{1}{K}\right]$, $D_p$ of MAPbI$_3$ and CdTe was calculated to be: MAPbI$_3$ - $D_p$ ($\Delta p$)=+1.2 or $D_p$ ($\Delta T$)=+2.6 ; CdTe - $D_p$ ($\Delta p$)= − 3.4 or $D_p$ ($\Delta T$)= −29.4. Despite the difference in the absolute value, the algebraic signs in both cases is usually consistent from both approaches (also for other materials that are presented Figure 1(i) - not presented here). Here we used only data for $\frac{\Delta E_g}{\Delta p}$, since $\frac{\Delta E_g}{\Delta T}$ include both thermal expansion effects (similar to pressure) and thermal (statistical distribution, vibrational) effects on $E_g$.

**Section B:**

## *About non-radiative recombination and the Huang-Rhys factor:*

**Fig. S8** shows two configuration coordinate diagrams that illustrate how electron-phonon coupling affects the efficiency of non-radiative recombination events. The y-axis of these diagrams shows the total (electronic + lattice) energy of the system. The upper parabola may represent the system with e.g. a free electron in the conduction band (with the conduction band edge at energy $E_2$) and the lower parabola that where the electron occupies a defect state (with energy $E_1$). The two parabolas are shifted relative to each other not only in energy but also in configuration coordinate (CC). This shift is a consequence of electron-lattice interaction and a stronger shift on the CC axis will typically make a non-radiative transition easier and therefore reduce the lifetime of the excited (higher energy) electron. To understand how a change in electron-lattice interaction will lead to faster recombination, we first have to study the process of electron capture in

**Fig. S8**(i). For the electron to be captured by the defect it needs to make a transition from the upper to the lower parabola. This can happen classically by thermal excitation to the crossing point (see

**Fig. S8**(i)) at energy $E_B$ above the conduction band edge or by tunneling to a vibrationally excited state of the lower parabola. Depending on the temperature, a different combination between thermal excitation and tunneling will be the most efficient way of electron transfer.[51,89] Let us now compare **Fig. S8**(i) with **Fig. S8**(ii). In the latter case, the shift on the CC axis is increased to a point where the minimum of the upper parabola intersects the lower parabola, i.e. the classical crossing point is at $E_B = 0$ and tunneling is not necessary. This is the worst-case scenario that leads to a peak of the transition rate as a function of the shift on the CC axis and that would allow fast recombination even at zero temperature. Thus, slow recombination requires the shift to be much less than what is shown in panel (ii). The shift of the parabola on the CC is typically measured in terms of either a Huang-Rhys factor $S_{HR}$ (more common in the inorganic semiconductor community) or in terms of a reorganization energy $\lambda = S_{HR}E_{ph}$ (common in the *molecular electron transfer and organic* semiconductor communities). **Fig. S8**(i) illustrates the definition of the Huang-Rhys factor. The value of the upper parabola at the position of the minimum of the lower parabola on the CC axis is defined as being $S_{HR}E_{ph}$ above $E_2$. The worst-case scenario in **Fig. S8**(ii) is reached if the energy difference between the two minima of the parabola, namely $E_0 = E_2 - E_1 = S_{HR}E_{ph}$. Thus for non-radiative recombination to be slow, the model requires $S_{HR}$ to be much smaller (or much larger, but that is not normally the case) than the number $p$ of phonons needed to dissipate the energy $E_0$. Thus, if the phonon energy is e.g. 30 meV, the difference between band edge and defect level is 600 meV, then $p = 20$ and $S_{HR}$ should ideally be substantially smaller than 20.

To better understand how the trends, developed in the current manuscript, affect non-radiative recombination, we use a more generic approach developed by Ridley[8,52] who provides analytical equations for $S_{\mathrm{HR}}$ that include the effect of either deformation coupling or polar coupling. For the case of deformation coupling, Ridley derives the Huang-Rhys factor of a generic defect as:

$$S_{\mathrm{HR}} = \frac{1}{2(\hbar\omega)^2} \frac{\hbar D^2}{M_{\mathrm{r}}\omega} I$$

and for polar coupling as:

$$S_{\mathrm{HR}} = \frac{3}{2(\hbar\omega)^2} \left\{ \frac{q^2(M_{\mathrm{r}}/V_0)\hbar\omega^2}{M_{\mathrm{r}}\omega q_{\mathrm{D}}^2} \left( \frac{1}{\varepsilon_\infty} - \frac{1}{\varepsilon_s} \right) \right\} I$$

These are the full expressions to what is presented in Eq. 24 and Eq. 25.

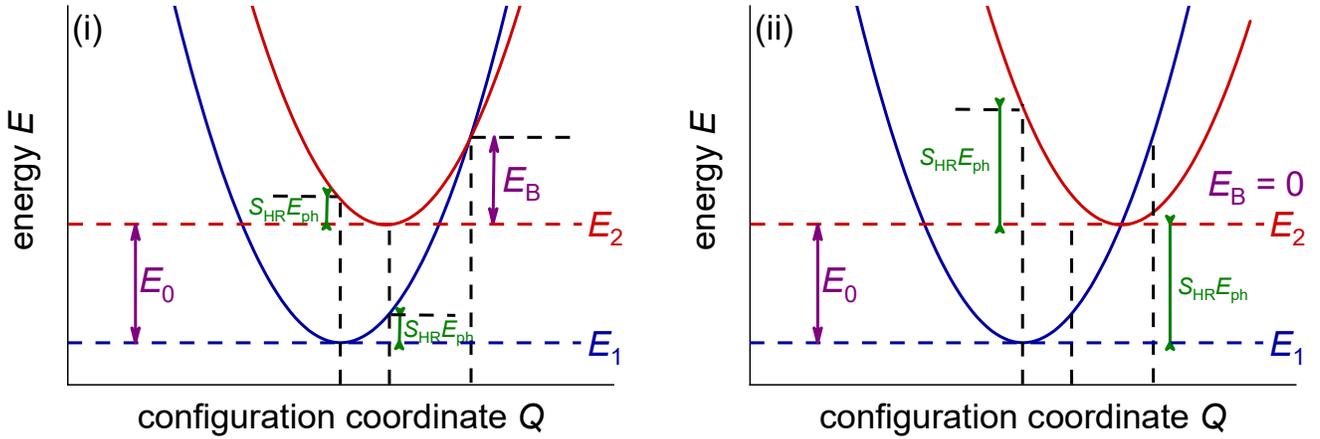

**Fig. S8**: Configuration coordinate diagrams, illustrating non-radiative transitions between two states (i.e. an electron in the conduction band and a defect state), represented by the upper and lower parabola. The two parabolas represent the total (electronic + vibrational) energy as a function of a single configuration coordinate.